%% file: ms.tex
\documentclass[12pt,a4paper]{article}
\usepackage[english]{babel}

\usepackage{amsmath}
\usepackage{amsfonts}
\usepackage{amssymb}
\usepackage{graphicx}
\usepackage[affil-it]{authblk}
\usepackage{hyperref}
\usepackage{color}
\usepackage{caption}
\usepackage{verbatim}
\usepackage[normalem]{ulem}
\usepackage{subfigure}
\usepackage{slashed}
\usepackage{cancel}
\usepackage{braket}
\usepackage{multicol}
\usepackage{tcolorbox}
\tcbuselibrary{theorems}
\usepackage{bm}
\date{}

\usepackage{enumitem}
\usepackage{float}

\newcommand{\micrOMEGAs}{\texttt{micrOMEGAs}}

\hypersetup{linktocpage=true}

\usepackage{tikz}
\usetikzlibrary{arrows,shapes}
\usetikzlibrary{trees}
\usetikzlibrary{matrix,arrows} 				
\usetikzlibrary{positioning}				
\usetikzlibrary{calc,through}				
\usetikzlibrary{decorations.pathreplacing}  
\usepackage{pgffor}							

\usetikzlibrary{decorations.pathmorphing}	
\usetikzlibrary{decorations.markings}
\tikzset{
    vector/.style={decorate, decoration={snake}, draw},
	provector/.style={decorate, decoration={snake,amplitude=2.5pt}, draw},
	antivector/.style={decorate, decoration={snake,amplitude=-2.5pt}, draw},
    fermion/.style={draw=black, postaction={decorate},
        decoration={markings,mark=at position .55 with {\arrow[draw=black]{>}}}},
    fermionbar/.style={draw=black, postaction={decorate},
        decoration={markings,mark=at position .55 with {\arrow[draw=black]{<}}}},
    fermionnoarrow/.style={draw=black},
    gluon/.style={decorate, draw=black,
        decoration={coil,amplitude=4pt, segment length=5pt}},
    scalar/.style={dashed,draw=black, postaction={decorate},
        decoration={markings,mark=at position .55 with {\arrow[draw=black]{>}}}},
    scalarbar/.style={dashed,draw=black, postaction={decorate},
        decoration={markings,mark=at position .55 with {\arrow[draw=black]{<}}}},
    scalarnoarrow/.style={dashed,draw=black},
    electron/.style={draw=black, postaction={decorate},
        decoration={markings,mark=at position .55 with {\arrow[draw=black]{>}}}},
	bigvector/.style={decorate, decoration={snake,amplitude=4pt}, draw},
}\usetikzlibrary{decorations.markings}

\tikzstyle{block} = [draw, rectangle, 
    minimum height=3em, minimum width=6em]

\title{Two Real Scalar WIMP Model in the Assisted Freeze-Out Scenario}
\author[1]{Basti\'an D\'iaz S\'aez\thanks{bastian.diaz@tum.de}}
\author[2]{Kilian Möhling \thanks{kilian.moehling@tu-dresden.de}}
\author[2]{Dominik Stöckinger \thanks{dominik.stoeckinger@tu-dresden.de}}
\affil[1]{Physik-Department, Technische Universität München, James-Franck-Straße, 85748 Garching, Germany}
\affil[2]{Institut für Kern- und Teilchenphysik, TU Dresden, Zellescher Weg 19, 01069 Dresden, Germany}

\begin{document}
\maketitle


\begin{abstract}
We study a simple dark matter model given by two interacting real singlet scalars, 
with only one of them coupled to the Higgs. The model therefore presents a minimal \textit{assisted}-\textit{freeze-out}
framework: both scalars contribute to the dark matter relic density, but only one of
them takes part in the elastic scattering with nuclei.
This reduces the expected interaction rate in direct detection experiments in such a way that the model in some regions of the parameter space may evade XENON1T constraints. We explore the model under theoretical (perturbativity, stability potential and unitarity) and experimental constraints (Higgs to invisible, relic density, direct detection), with the use of the narrow-width-approximation near the Higgs resonance to calculate the averaged annihilation cross section. We show that the model is viable in the Higgs resonance region and for scalar singlets masses of hundreds of GeV, although the model becomes highly constrained at high energies.
\end{abstract}


\newpage
\tableofcontents


\input{0-Introduction_new}

\input{1-Tworeals_new}

\appendix
\input{appendix}

\input{appendixb}


\bibliography{bibliography}
\bibliographystyle{utphys}

\end{document}

%% file: 0-Introduction_new.tex
\section{Introduction}
While cosmological observations over the past years have made it more and more
apparent that Dark Matter (DM) is an important part of our universe
\cite{planck2018}, direct detection experiments have put strong constraints on
the parameter space of many WIMP models, in particular one of the most popular 
DM models: the real singlet scalar or also called \textit{Singlet Higgs Portal} (SHP) \cite{Cline:2013gha, Athron:2017kgt}. 
Even when considering the singlet scalar as just a partial DM source together with
additional non-thermal contributions (e.g. axions 
\cite{Dasgupta:2013cwa}), most of the parameter space around a few hundred GeV up to the TeV scale
is ruled-out.
Following this minimal approach, models with two real singlet scalars have been studied
in the last years, reopening the possibility 
of having scalar DM near the weak scale. 
Two classes of models have been explored: one scalar stabilized by a $Z_2$
symmetry with the other scalar being unstable \cite{Abada:2011qb, Ahriche:2013vqa,Ghorbani:2014gka,
Casas:2017jjg, Shajiee:2018jdq, ARHRIB201915, Maity:2019hre, Maniatis:2020ois} 
or both scalars stabilized by two symmetries $Z_2\times Z_2'$ \cite{Maity:2019hre,
Belanger:2011ww, Bazzocchi:2012pp, Modak:2013jya, Ghorbani:2015xvz, Bhattacharya:2016ysw, Pandey:2017quk, Ghorbani:2019itr}
\footnote{Fermions and vectors have also been considered in similar frameworks, e.g.
\cite{Bian:2013wna, Bhattacharya:2013hva, Agashe:2014yua}.}. 
In the latter case, when the two scalars are WIMPS with sizable interaction between them \cite{Bhattacharya:2016ysw}, the model may give the correct relic density on large portions of the parameter space, but most of it is excluded by XENON1T direct detection results. Only  
the Higgs resonance region, where a more careful treatment of the thermally averaged 
annihilation cross section is necessary, seems to remain viable. In view of this, it may be of interest to look for regions of parameter space in this framework which evade the strong direct detection constraints, possibly allowing DM with masses of several hundred GeV up to a few TeV. For instance, in \cite{Maity:2019hre} (also see \cite{Arcadi:2016qoz, DiazSaez:2021pmg}) such a two-component scalar model is explored under the assumption that one of the scalars couples very weakly to the Higgs ($\lambda \sim 10^{-5}$), giving rise to a chronology in the decoupling process (\textit{exchange driven freeze-out}), with DM conversion being a pivotal process in the determination of the relic abundance. They show that this mechanism successfully evades XENON1T bounds, allowing DM with masses of hundreds of GeV, provided certain quartic coupling take values above the unity.


In this work we study the two real scalar model, both stable gauge singlets,
with one of the Higgs portals set to zero. In this regime the model becomes a minimal realization of an appealing idea, the so-called assisted freeze-out mechanism \cite{Belanger:2011ww} (for a previous similar idea see \cite{Cao:2007fy}). The motivation of this approach is that only one of the scalars interacts directly with baryonic matter through its Higgs portal, whereas the other scalar may contribute most to the DM budget. In this way, the Higgs coupled scalar may evade the direct detection constraints more easily due to
the rescaling of the expected DM direct detection rate by its partial abundance: $\Omega_{coupled}/\Omega_{total}$. Qualitatively, this framework can be viewed as an approximation of the exchange driven freeze-out \cite{Maity:2019hre}.  

We study the parameter space in detail to identify viable coupling and mass ranges of the assisted freeze-out framework. For masses above the Higgs resonance, we find that large couplings $\lambda\gtrsim1$ are required in order to pass both direct detection and relic density constraints, in agreement with \cite{Maity:2019hre}. We then further impose theoretical constraints on the resulting parameter space in order to see to what extent the resulting high couplings are consistent with perturbative, scalar potential stability and unitarity constraints. We also specifically study the resonance effects for DM masses around 60 GeV, half of the Higgs boson mass, where the real singlet scalar in SHP is not yet ruled out. Finally, we study the effects of the renormalization group equations on the remaining parameter space after applying theoretical and experimental constraints.

In the following section, we present the model along with the theoretical constraints. In section 3, we present the corresponding Boltzmann equations,
approximations to the averaged annihilation cross section and we study the predicted relic 
abundance in detail, specifically in comparison with the SHP. 
In section 4, we show some scan results inside and outside the resonance region, taking 
into account the various constraints including those given by Landau poles of the model. Finally, we give a brief summary and the
conclusions in section 5.

%% file: 1-Tworeals_new.tex
\section{Model and Theoretical Constraints}\label{modelandtheoconst}
The model introduces two real gauge singlet scalars $S_1$ and $S_2$, each one stabilized by $
\mathbb{Z}_2\times\mathbb{Z'}_2$ symmetry, with the fields transforming as:
	\begin{subequations}\label{eq:two_reals_trans_under_Z2xZ2p}
	\begin{align}
  		\mathbb{Z}_2&\colon S_1\to -S_1,\ S_2\to S_2,\ \text{SM}\to\text{SM}, \\
  		\mathbb{Z}'_2&\colon S_1\to S_1,\ S_2\to -S_2,\ \text{SM}\to\text{SM}.
	\end{align}
	\end{subequations}
Consequently, the scalar potential for this model is
	\begin{equation}\label{pot1}
  	V(H,S_1,S_2)=\mu^2_H|H|^2+\lambda_H|H|^4+\sum_{i=1,2}\left(\frac{\mu^2_i}{2}S^2_i
  	+\frac{\lambda_i}{4!}S^4_i + \frac{\lambda_{Hi}}{2}|H|^2S^2_i\right)
  	+\frac{\lambda_{12}}{4}S^2_1S^2_2 .
	\end{equation}
The tree-level scalar masses after EWSB are
	\begin{align} \label{masses}
	m_h^2 &= 2v_H^2\lambda_H, \\
	m_i^2 &= \mu_i^2 + \dfrac{1}{2}v_H^2\lambda_{Hi},
	\end{align}
with $v_H$ the Higgs vacuum expectation value of 246 GeV. In order for the singlet scalars to not acquire a vacuum expectation value, we set $\mu_i^2 > 0$, then the physical scalar states are: $\{h, S_1, S_2\}$. 

According to the normalization of our scalar potential eq. \eqref{pot1}, the stability of the potential at tree-level requires \cite{Kannike:2012pe, Kannike:2016fmd, Ferreira:2016tcu}
	\begin{eqnarray}\label{pot}
	\lambda_H>0,\quad\lambda_1 > 0,\quad\lambda_2>0,\quad\lambda_a>0,\quad\lambda_b > 0,\quad\lambda_c>0 , \\
	\sqrt{2\lambda_H\lambda_1\lambda_2/9} + \lambda_{H1}\sqrt{\lambda_2/3} + \lambda_{H2}\sqrt{\lambda_1/3} + \sqrt{2\lambda_a\lambda_b\lambda_c} > 0,
	\end{eqnarray}
where $\lambda_a \equiv \lambda_{H1} + \sqrt{2\lambda_H\lambda_1/3}$, $\lambda_b \equiv \lambda_{H2} + \sqrt{2\lambda_H\lambda_2/3}$ and $\lambda_c \equiv \lambda_{12} + \sqrt{\lambda_1\lambda_2/9}$\footnote{From these last constraints we obtain that, for instance, if $\lambda_{H1}$ is negative, it must to satisfy $\lambda_{H1}^2 < 2\lambda_H\lambda_2/3$, and if it is positive, then $\lambda_{H1} > 0$.}. Taking $m_h =$ 125 GeV in eq. \eqref{masses} gives $\lambda_H = 0.13$. 

On the other hand, tree-level perturbative unitarity puts constraints on the zeroth partial wave of the system \cite{Lee:1977eg, Cynolter:2004cq}, which for a $2\rightarrow 2$ elastic scattering is given by
\begin{eqnarray}
 a_0 = \frac{1}{32\pi}\sqrt{\frac{4|\vec{p}_f||\vec{p}_f|}{s}}\int_{-1}^{+1}\mathcal{M}_{2\rightarrow 2}d\cos\theta ,
\end{eqnarray}
with $s$ the center-of-mass frame (CM) energy and $p_{i,f}^{\text{CM}}$ the tree-momenta of initial and final particles in the CM. Unitarity then requires $|\text{Re}(a_0)| \leq \frac{1}{2}$. Considering the six-channel system 
\begin{eqnarray}
 \{hh, S_1S_1, S_2S_2, S_1S_2, hS_1, hS_2\},
\end{eqnarray} 
the zero-partial wave matrix at high energies $s\gg m_h, m_1, m_2$ is
\begin{eqnarray}
a_0\rightarrow \frac{1}{16\pi}
\begin{pmatrix}
3\lambda_H & \lambda_{H1} & \lambda_{H2}& 0 & 0 & 0\\
\lambda_{H1} & \lambda_{1} & \lambda_{12} & 0 & 0 & 0\\
\lambda_{H2} & \lambda_{12} & \lambda_{2} & 0 & 0 & 0\\
0 & 0 & 0 & \lambda_{12} & 0 & 0\\
0 & 0 & 0 & 0 & \lambda_{H1} & 0\\
0 & 0 & 0 & 0 & 0 & \lambda_{H2}\\
\end{pmatrix}.
\end{eqnarray}
Perturbative unitarity constraints on the matrix $a_0$ result in bounds on its eigenvalues $e_i$. In this work we concentrate on the case in which $\lambda_{H2} = 0$, which then results in the following constraint on the eigenvalues of the $3\times 3$ upper block
\begin{eqnarray}\label{eigen}
 |e_i (\lambda_1, \lambda_2, \lambda_{H1}, \lambda_{12})| \leq 8\pi , \quad i=1,2,3.
\end{eqnarray}
For simplicity we omit the analytical expressions for each eigenvalue. 

Another theoretical constraint is given by perturbativity, which implies approximate upper limits for all coupling in order the perturbative expansion be reliable. At this point we do not choose any specific value for these upper limit. In the following discussions we take into account all three theoretical constraints (stability, unitarity and perturbativity) and analyze their impact on the viable parameter space.
\begin{figure}[t!]
\centering
\includegraphics[width=0.3\textwidth]{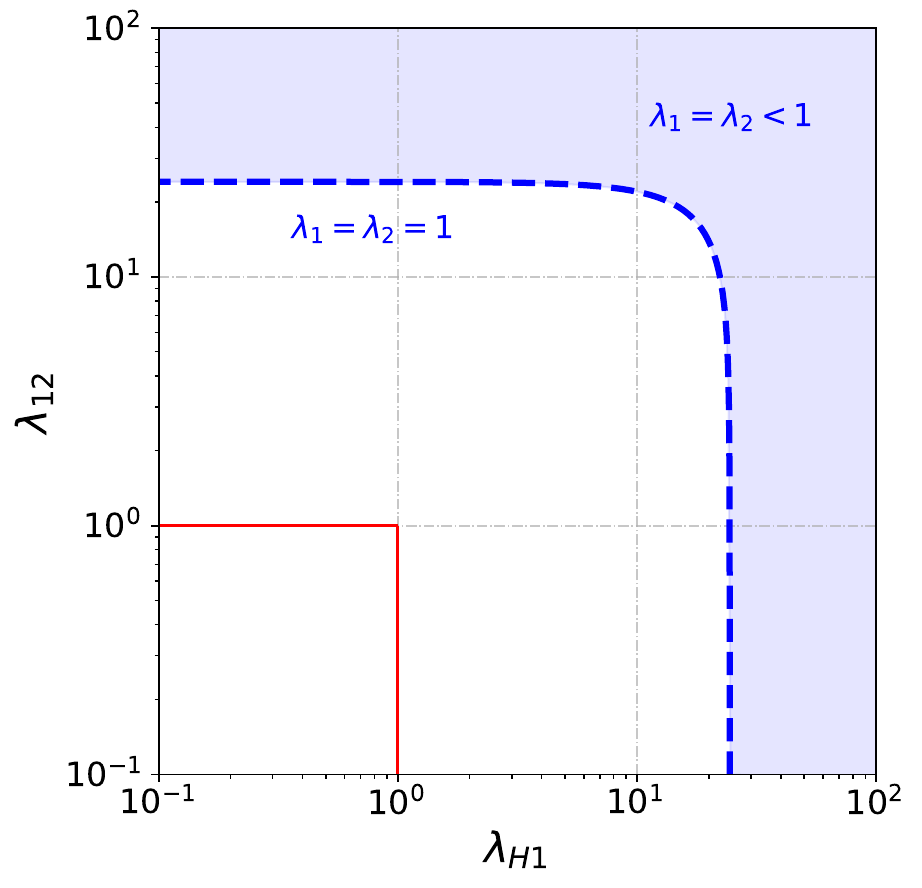}
\includegraphics[width=0.3\textwidth]{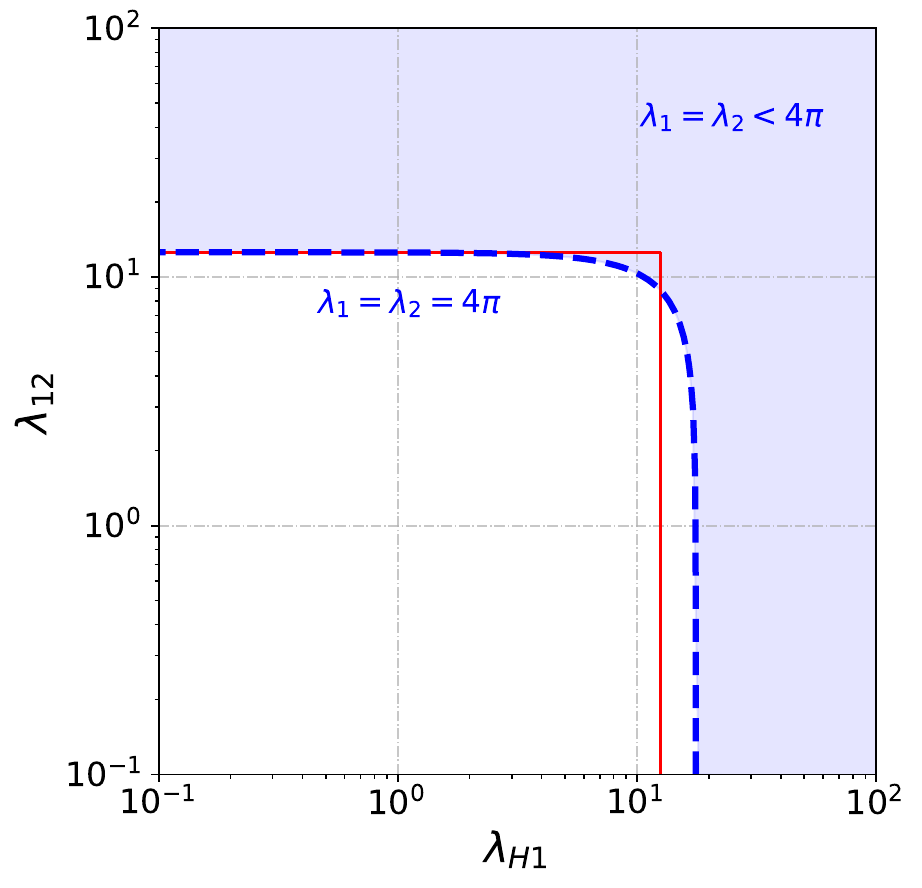}
\includegraphics[width=0.3\textwidth]{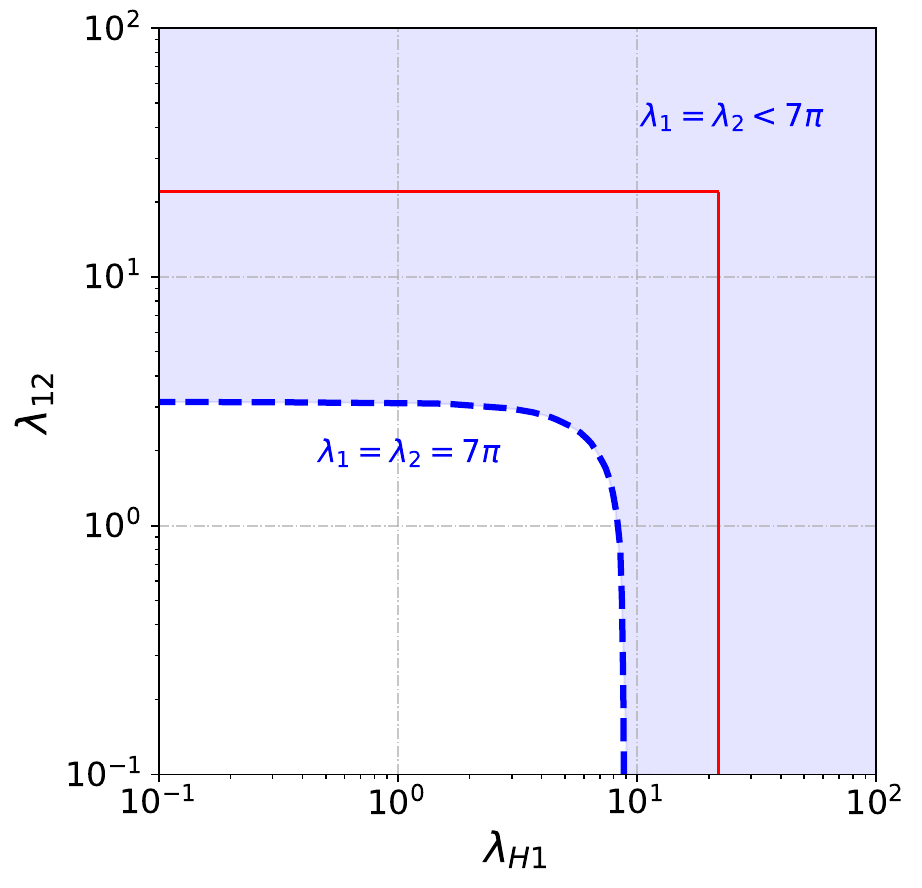}
\caption{Unitarity constraints on the pair $(\lambda_{H1},\lambda_{12})$ for $\lambda_{1,2} = 1, 4\pi, 7\pi$, respectively, with the blue region excluded by this constraint. The region inside of each red square may be interpreted as the allowed parameter space by perturbativity when the perturbative limit on the couplings is $1, 4\pi, 7\pi$, respectively.}
\label{fig:theo}
\end{figure}

Considering only positive values for $\lambda_{H1}$ and $\lambda_{12}$ (negative values do not change the analysis of our work) then the stability potential constraints eq. \eqref{pot} do not give any new information. In Fig.~\ref{fig:theo} we show the unitarity bounds on $\lambda_{H1}$ and $\lambda_{12}$ for the choices $\lambda_{1,2} = 1, 4\pi$ and $7\pi$, respectively, with the blue region excluded by this constraint. The bigger the values of the quartic couplings $\lambda_{1,2}$ the stronger are the constraints on the remaining pair $\lambda_{H1}, \lambda_{12}$. In particular, as the middle plot shows, it is not possible to have the quartic couplings $\lambda_{1,2}$ and $\lambda_{H1},\lambda_{12}$ all simultaneously with values above $4\pi$. Additionally, the region inside the red squares indicates the regions allowed by different choices of the perturbativity limit. The limits are set to $1$, $4\pi$, and $7\pi$, respectively, i.e.\ to the same values as the quartic couplings in each plot. Clearly, perturbativity places limits on each individual coupling, while unitarity places limits on combinations of couplings. If very strong perturbativity limits are imposed as in the left plot, unitarity does not imply additional constraints, while very loose perturbativity limits, as in the right panel, are much less stringent than unitarity limits.


Therefore, the analysis shows that the  theoretical constraints depend on both the assumed values for the quartic couplings $\lambda_{1,2}$ as well as the choice of the perturbativity limit. In the last part of this work, when we contrast theoretical and experimental constraints on the model, we discuss the viability of parameter space depending on the chosen criteria.

\section{Relic Abundance}
In this section we discuss the main formulas describing the evolutions of the DM number density 
throughout the thermal history. We introduce precise approximations for the averaged 
annihilation cross section and study the behavior of the DM relic density both in and outside
the Higgs resonance region. We also point out that a careful treatment of the averaged
annihilation cross section near the Higgs resonance is necessary in order to obtain the
correct results.
\subsection{Boltzmann equations}\label{beqsub}
The coupled Boltzmann equations for the DM number densities of $S_1$ and $S_2$ are given by
	\begin{align}\label{beq_num_den}
	\begin{split}
	\dfrac{dn_1}{dt} + 3Hn_1 = &-\braket{\sigma_{11XX} v}\Bigl[n_1^2 - 
	n_{1e}^2\Bigr] - \braket{\sigma_{1122} v}\Bigl[n_1^2 - 
	\dfrac{n_{1e}^2}{n_{2e}^2}n_2^2\Bigr]\Theta(m_1 - m_2) \\
	&+\braket{\sigma_{2211} v}\Bigl[n_2^2 - \dfrac{n_{2e}^2}{n_{1e}^2}
	n_1^2\Bigr]\Theta(m_2 - m_1), \\
	\dfrac{dn_2}{dt} + 3Hn_2 = &-\braket{\sigma_{22XX} v}\Bigl[n_2^2 - 
	n_{2e}^2\Bigr] - \braket{\sigma_{2211} v}\Bigl[n_2^2 - 
	\dfrac{n_{2e}^2}{n_{1e}^2}n_1^2\Bigr]\Theta(m_2 - m_1) \\
	&+\braket{\sigma_{1122} v}\Bigl[n_1^2 - \dfrac{n_{1e}^2}{n_{2e}^2}
	n_2^2\Bigr]\Theta(m_1 - m_2),
	\end{split}
	\end{align}
with $X$ being some SM particle, $\braket{\sigma v}$ the thermal average of the cross section times the relative velocity of a given process $ii\rightarrow jj$, with $i,j=1,2,X$, and, assuming a Maxwell-Boltzmann thermal distribution, the equilibrium number densities $n_{ie}$ as a function of the thermal bath temperature
$T$ are given by
	\begin{align}\label{eqden}
	n_{ie}(T) = g_i\dfrac{m_i^2}{2\pi^2}TK_2(\tfrac{m_i}{T}),
	\end{align}
with the internal degrees of freedom $g_i$, and the modified Bessel function of the second kind 
$K_2$. Furthermore, the Hubble parameter $H$ and entropy density $s$ in the early universe can
be written as
	\begin{align}
	H(T)=\sqrt{\dfrac{4\pi^3G}{45}g_{*}(T)} \cdot T^2, \quad 
	s(T)=\dfrac{2\pi^2}{45}g_{*s}(T)\cdot T^3 ,
	\end{align}
with $G$ is the Newton gravitational constant, and $g_{*}(T)$ and $g_{*s}(T)$ are the effective degrees of freedom contributing respectively to the energy and the entropy density at temperature $T$. For our numerical evaluations we interpolate $g_{*}$ and $g_{*s}$ from the values given in
\cite{Husdal:2016haj}.
We have introduced Heaviside functions in eq. \eqref{beq_num_den} in order to select the 
formulation of the interaction term in which $n_{ie}^2/n_{je}^2\rightarrow 0$ at small $T$,
making the numerical evaluation more convenient.

Lastly, introducing $x := \mu / T$, where $\mu = m_1m_2/(m_1 + m_2)$ and $Y_i := n_i/s$, we 
rewrite eq. \eqref{beq_num_den} in the usual dimensionless form
	\begin{align}\label{beq_yield}
	\begin{split}
	\dfrac{dY_1}{dx} = &-\lambda_{11XX}\cdot[Y_1^2-Y_{1e}^2]-\lambda_{1122}\cdot
	[Y_1^2-r_{12}^2Y_2^2]\cdot\Theta(m_1-m_2)\\
	&+\lambda_{2211}\cdot[Y_2^2-r_{21}^2Y_1^2]\cdot\Theta(m_2-m_1), \\
	\dfrac{dY_2}{dx} = &-\lambda_{22XX}\cdot[Y_2^2-Y_{2e}^2]-\lambda_{2211}\cdot
	[Y_2^2-r_{21}^2Y_1^2]\cdot\Theta(m_2-m_1)\\
	&+\lambda_{1122}\cdot[Y_1^2-r_{12}^2Y_2^2]\cdot\Theta(m_1-m_2),
	\end{split}
	\end{align}
with
	\begin{align}
	r_{ij}(x):=\dfrac{Y_{ie}(x)}{Y_{je}(x)},\quad \lambda_{iijj}(x):=
	\dfrac{\langle\sigma_{iijj}v\rangle(x)\cdot s(T)}
	{x\cdot H(T)} \qquad \text{for} ~ i,j=1,2,X.
	\label{def_lambda}
	\end{align}
The thermal average annihilation cross section is given by
\begin{align}\label{mainaverage}
	\langle \sigma_{iijj} v \rangle (x) = \int_{4m_i^2}^{\infty}
	\dfrac{s \sqrt{s-4m_i^2} K_1(\sqrt{s}\tfrac{x}{\mu})(\sigma v)_{iijj}}
	{16\frac{\mu}{x}m_i^4 K_2^2(m_i \tfrac{x}{\mu})} ds,
\end{align}
where $K_1$ and $K_2$ are modified Bessel functions of the second kind. 
Properly evaluating this expression plays a significant role in
determining the DM relic density, especially near resonances and thresholds \footnote{The expression \eqref{mainaverage} is based on the assumption of local thermal equilibrium when chemical decoupling is taking place. However, near a resonance, this assumption may be no longer valid, then a more involved treatment must to be carried \cite{Binder:2017rgn}. We will comment more about this point in Sec.~\ref{scanresults}.}. In order to evaluate eq. \eqref{mainaverage}, we use the narrow width approximation (NWA) in the resonance region \cite{Gondolo:1990dk, Feng:2017drg}, whereas outside of it we took the constant approximation $\langle\sigma v\rangle = \sigma v\left(s = 4m_i^2\right)$. We used the perturbative tree-level $\sigma v$ into kinematically open channels, with the total decay width of the Higgs boson $\Gamma_h = 1.34\times 10^{-3}$ MeV, supplemented with the invisible decay width when $m_1 \leq m_h/2$. These cross sections are in agreement with the analytic results obtained in \cite{Bhattacharya:2016ysw}. We solved the Boltzmann equations in eq.~\eqref{beq_yield} numerically using \textit{solve-ivp} from Scipy module in Python. Our results have been cross-checked with MicrOMEGAS.

Finally, after solving the Boltzmann equations the DM density parameter is given by 
	\begin{align}\label{relic}
	\Omega h^2 = \Omega_1 h^2 + \Omega_2 h^2, ~~\qquad
	\Omega_i h^2 = \dfrac{2.9713\cdot 10^9}{10.5115 ~\text{GeV}}
	\cdot Y_{i,0} \cdot m_i ,
	\end{align}
where $Y_{i,0}$ is the yield of $S_i$ today, i.e. after freeze-out.

\subsection{Assisted freeze-out}\label{sec:assisted_freeze-out}
We now focus on the assisted freeze-out scenario, i.e. we set one Higgs portal coupling to zero, $\lambda_{H2}=0$\footnote{In Appx.~\ref{oneloop} we discuss about the effects of radiative corrections of the renormalized value of $\lambda_{H2}$ on the relic abundance calculation. The effects of radiative $\lambda_{H2}$ becomes non-negligible only in those cases in which the combination $\lambda_{H1}\lambda_{12}$ takes very high values, then in this way deviating the predictions of the relic density values found in the rest of the paper (i.e. $\lambda_{H2} = 0$) at most in a factor of two. For studies considering a non-vanishing tree-level value $\lambda_{H2}$ see \cite{Bhattacharya:2016ysw}.}. In this way, from now on only $S_1$ will be \textit{coupled} to the SM, whereas $S_2$ becomes \textit{decoupled}. Consequently, $\lambda_{22XX} = 0$ in eq.~(\ref{beq_yield}) and the relevant parameters of this scenario are $m_1, m_2, \lambda_{H1}$ and $\lambda_{12}$. We allow sizable couplings $\lambda_{H1}$ and $\lambda_{12}$, such that the coupled scalar starts in thermal equilibrium with the SM, and at the same time brings the decoupled one into the equilibrium too. 
In the second part of this section we exemplify the assisted freeze-out mechanism, showing clear differences between the predictions of the real singlet scalar $S$ in the \textit{Singlet Higgs Portal} model (SHP) \cite{Cline:2013gha} and the equivalent $S_1$ scalar in this scenario.

\begin{figure}[h!]
\centering
\includegraphics[width=.5\textwidth, trim=10 0 40 20, clip]{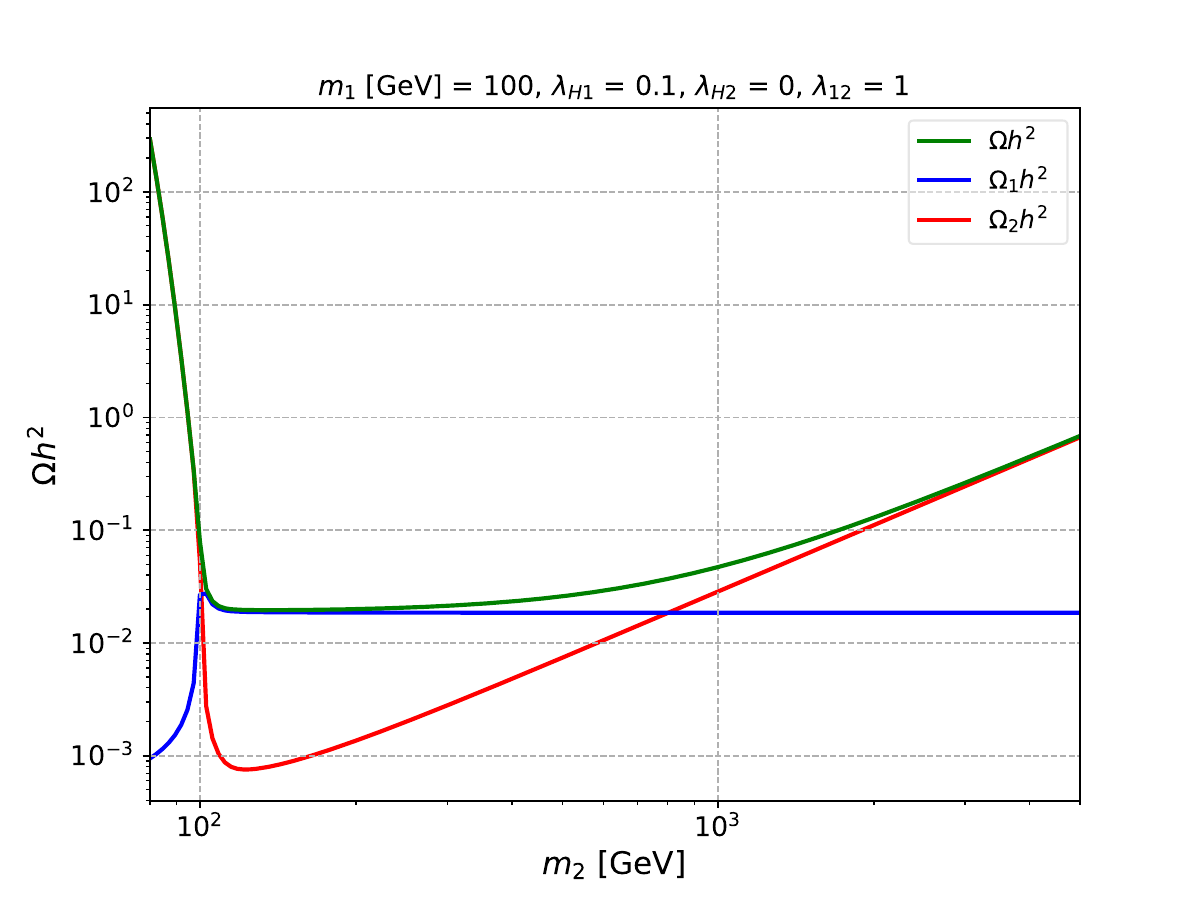}\hspace*{-2mm}
\includegraphics[width=.5\textwidth, trim=10 0 40 20, clip]{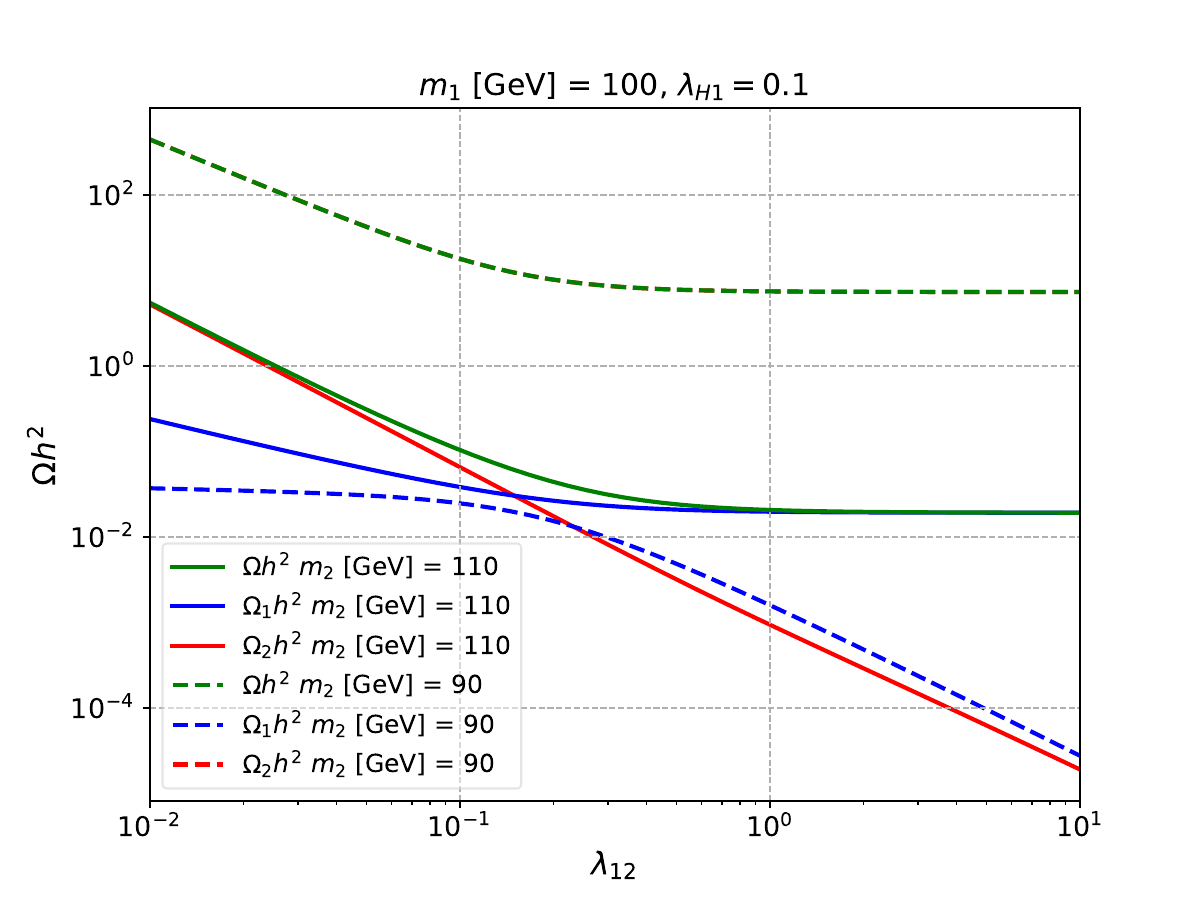}
\caption{DM density parameters $\Omega h^2$ (green), $\Omega_1 h^2$ (blue) and $\Omega_2 h^2$ (red) as functions of
$m_2$ (left) and $\lambda_{12}$ (right), with $m_1=100$ GeV and $\lambda_{H1}=0.1$ in both plots. In the left plot we 
have set $\lambda_{12}=1$, and in the right plot $m_2=90$ GeV (dashed) and $m_2=110$ GeV (solid).}
\label{fig:omega_dependence}
\end{figure}

The dynamics of the abundance evolution is highly dependent on the mass hierarchy of the two singlet scalars. The behaviour of the resulting relic densities can be seen in more detail in Fig.~\ref{fig:omega_dependence} (left), which fixes $m_1 = 100$ GeV and varies $m_2$. For $m_2 < m_1$ (inverted hierarchy) the 
overwhelmingly dominant contribution to the total relic comes from $\Omega_2$, while in the normal mass hierarchy ($m_2 > m_1$) $\Omega\approx\Omega_2$ only at very high mass 
differences with a much smaller relative difference between $\Omega_1$ and $\Omega_2$. 
Again, this behaviour can be understood by looking at the corresponding Boltzmann Equation.
In the normal hierarchy we already mentioned that if $m_2 \gg m_1$, we expect $Y_2$ to evolve independently of $Y_1$.
In fact, since $\langle\sigma_{2211} v\rangle\sim m_2^{-2}$, eq. \eqref{beq_yield} effectively
decouples, resulting in $Y_1\sim const.$ (i.e. $\Omega_1 \sim const.$) and $Y_2\sim m_2^2$ ($\Omega_2 \sim m_2^3$), as it is shown in Fig.~\ref{fig:omega_dependence} (left) with the blue and red lines, respectively. Although in the normal hierarchy at high mass differences we have
$\Omega_2 \gg \Omega_1$, we typically obtain a too small overall DM relic abundance. This suggests that we would have to go to smaller couplings $\lambda_{H1}$, 
which in turn will produce larger relative contribution of $\Omega_1$ to $\Omega$.
In the inverted hierarchy,
constraints on the DM relic abundance also put very strong 
constraints on the possible masses $m_2$ for given $m_1, \lambda_{H1}$ and $\lambda_{12}$.
This will be an important point to keep in mind when choosing the mass resolution in
parameter space scans.

Considering the $\lambda_{12}$ dependence of the DM relic abundance, we would 
expect a larger coupling to lead to a later freeze out and therefore lower relic densities.
For small couplings $\lambda_{12}\sim\lambda_{H1}$ this behavior can indeed be seen in
Fig. \ref{fig:omega_dependence} (right),
however, at large couplings $\lambda_{12}\gg\lambda_{H1}$, $\Omega$ becomes asymptotically
constant. In the normal mass hierarchy this is again explained by the fact that
$dY_2/dx\sim -\lambda_{2211}Y_2^2$, and therefore $Y_2\sim 1/\lambda_{12}^2$. Since the
interaction term is subsequently roughly proportional to 
$\lambda_{12}^2Y_2^2\sim 1/\lambda_{12}^2$, we find $Y_1\sim const.$ at large 
couplings. Interestingly, in the inverted mass hierarchy the total relic abundance behaves 
very much in the same way, although the independence of $\Omega$ on $\lambda_{12}$ at large 
couplings now comes from the fact that $Y_2\sim const$. The explanation is very similar to 
that in the normal hierarchy: we have $dY_1/dx\sim -\lambda_{1122}Y_1^2$, i.e. 
$Y_1\sim1/\lambda_{12}^2$ and therefore $Y_2\sim const$. 

\begin{figure}[t!]
\centering
\includegraphics[width=0.5\textwidth]{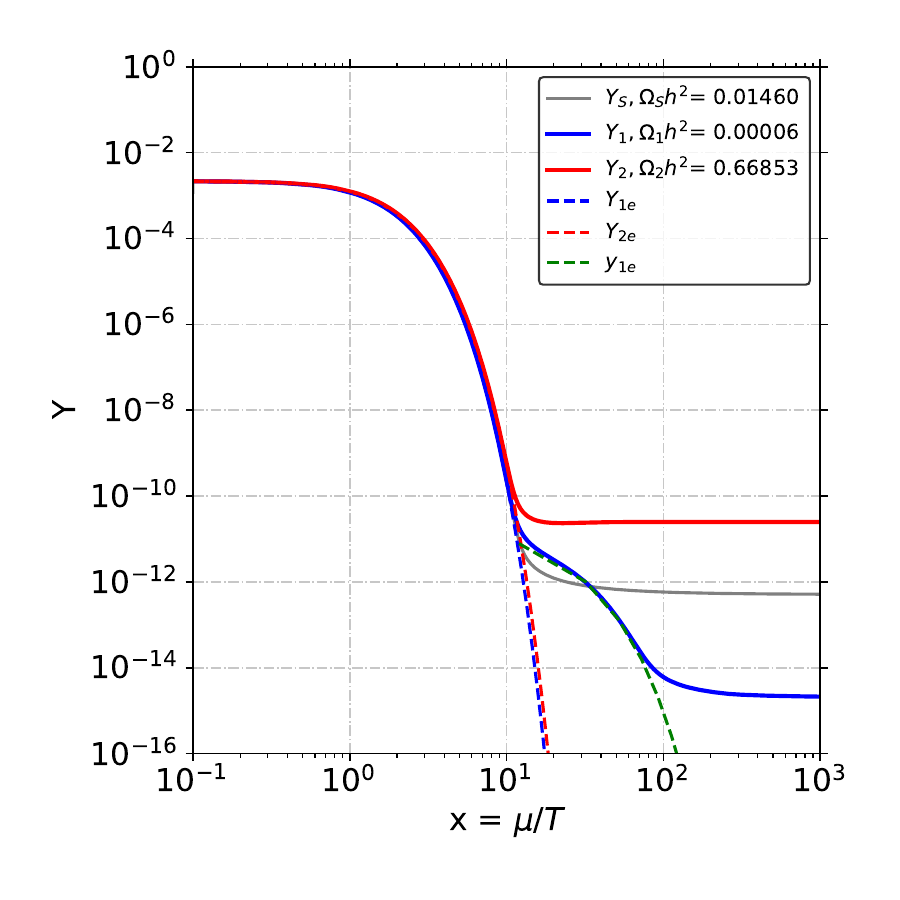}
\caption{Thermal evolution of the abundances in the assisted freeze-out scenario as a function of $x$, for $m_1 = 100$ GeV, $m_2 = 95$ GeV and $(\lambda_{H1}, \lambda_{H2}, \lambda_{12}) = (0.1, 0 , 10)$. The blue (red) solid line corresponds to yield of $S_1$ ($S_2$), and the gray one the yield of the singlet scalar in the SHP. The dashed blue and red lines correspond to the equilibrium densities of $S_1$ and $S_2$, respectively, whereas the green dashed line corresponds to the new equilibrium $y_{1e}$ that $S_1$ follows after it decouples from the SM bath.}
\label{fig:evo_goal}
\end{figure}

Having made a detailed analysis of the relic abundances as a function of the parameters, here we exemplify the mechanism operating in the inverted hierarchy that will be useful in the rest of the paper. In Fig.~\ref{fig:evo_goal} we show the evolution of $Y_1$ (solid blue line) and $Y_2$ (solid red line) as a function of $x$ when the assisted mechanism is present, with the dashed blue and red lines corresponding to the equilibrium densities of $S_1$ and $S_2$, respectively. Right after $S_1$ decouples from the SM thermal bath, it enters into a new equilibrium $y_{1e}$ (green dashed line), where annihilations $S_1S_1\rightarrow S_2S_2$ continue being effective, decreasing $Y_1$ much more than in the singlet scalar in the Singlet Higgs Portal (SHP) (grey solid curve). The modified equilibrium function is given by $\braket{\sigma_{1122} v}(Y_{1e}/Y_{2e})^2Y_2^2$, obtained directly from the Boltzmann equation in eq.~\eqref{beq_yield}. This behavior is a generic feature of interacting multicomponent DM \cite{Bhattacharya:2013hva, Bhattacharya:2016ysw}, which we will utilize to evade direct detection constraints. Finally, this effect is present independent of the value of $m_1$, but it depends strongly on the singlet mass difference and $\lambda_{12}$, as exemplified in Fig.~\ref{fig:omega_dependence}.


\section{Dark Matter Phenomenology}\label{dmpheno}
In this section, we start reviewing relevant experimental constraints, such as Higgs decay to invisible at the LHC, relic density and direct detection, to subsequently applied them onto the promising parameter space of the model. Similarly to the last section, we focus on the Higgs resonance region separately and contrast the results with the SHP predictions. Finally, we present some scans on the parameter space outside the Higgs resonance, contrasting those points that fulfill the experimental constraints with theoretical bounds. 
\subsection{Experimental Constraints}\label{sec:exp_const}
The strongest constraint on the parameter space is given by the observed DM relic abundance.
We use the value given by the most recent Planck data \cite{planck2018}:
	\begin{align}
	\Omega_ch^2 = 0.11933 \pm 0.00091, \qquad \text{at } 2\sigma .
	\end{align}
We will denote this measured value of the DM relic by $\Omega_c$ and the DM relic predicted 
by the model by $\Omega$. This gives a parameter space constraint $\Omega=\Omega_c$, which
we applied with a tolerance of $\pm 5\%$ accounting for uncertainties in our computation. 

The second constraint we need to pass is given by direct detection experiments such as
XENON1T. These give an upper bound on the spin-independent (SI) 
DM nucleon scattering cross section $\sigma_{SI}$, 
	\begin{align}
	\sigma_{SI,i} = \dfrac{\lambda_{Hi}^2f_N^2}{4\pi}\dfrac{\mu_i^2m_n^2}{m_h^4m_i^2}, \quad i=1,2,
	\end{align}
with $f_N = 0.3$ the effective Higgs-nucleon coupling, $m_n = 0.9$ GeV the nucleon mass and
$\mu_i=m_nm_i/(m_n+m_i)$ the DM-nucleon reduced mass 
\cite{Cline:2013gha, Athron:2017kgt}.
Since in our case only one of the scalars interacts with the SM, i.e. $S_1$, the differential 
detection rate is 
$dR/dE\propto \sigma_{SI}n_{1\odot}=\sigma_{SI}(\Omega_1/\Omega)n_{\odot}$, where 
$n_{\odot}$ denotes the local DM density. This effectively rescales the direct detection
constraint on our model to $(\Omega_1/\Omega)\sigma_{SI,1}\leq \sigma_{Xe}$, where
$\sigma_{Xe}$ is the upper bound given by XENON1T \cite{Aprile_2018}.
In the SHP, it has been shown that indirect bounds set much weaker constraints on the
parameter space than the direct detection limits \cite{Cline:2013gha}. In our scenario, this 
still holds, since the only parameter space passing the stringent XENON1T constraints is in
the inverted mass hierarchy where $\Omega_1/\Omega\ll 1$, therefore primary fluxes produced
from the annihilation of a pair of $S_1$ are highly suppressed.

Lastly, for $m_1 \leq m_h/2$, the Higgs boson can decay into two $S_1$, with an invisible
decay width given by
	\begin{align}
	\Gamma_{inv}(h\to S_1S_1) = \frac{\lambda^2_{H1}v_H^2}{32\pi m_h}
	\sqrt{1 - \frac{4m^2_1}{m^2_h}}.
	\label{h11}
	\end{align}
This contributes to the invisible branching rate 
$\text{Br}(h\to\text{inv})=\Gamma_\text{inv}/(\Gamma_\text{SM}+\Gamma_\text{inv})$, with 
$\Gamma_\text{SM}= 4.07$ MeV \cite{Sirunyan:2018owy}. Experimental searches put strong 
constraints on this quantity, with the most stringent value being given by 
$\text{Br}(h\to\text{inv}) < 0.19$ at $95\%$ C.L. \cite{Sirunyan:2018owy}.
 
\subsection{Scan Results}\label{scanresults}
\begin{figure}[h]
\centering
\includegraphics[width=0.45\textwidth, trim=0 10 50 50, clip]{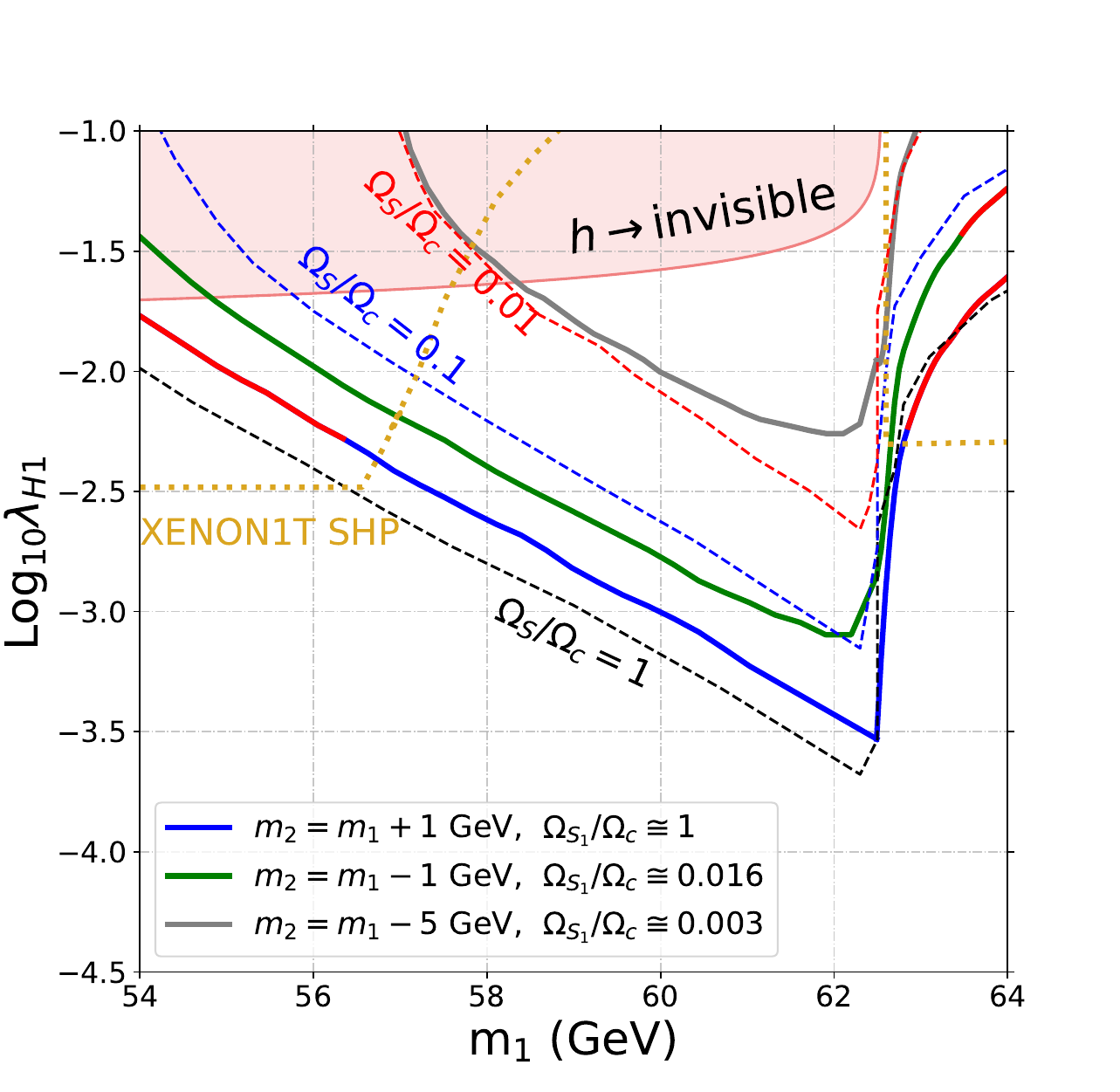}\quad
\caption{Contours of fixed relic abundance in the assisted freeze-out model (solid) 
and in the SHP (dashed) in the parameter space $(m_1,\lambda_{H1})$ around the Higgs 
resonance. 
All contours of the assisted freeze-out model fulfill
$\Omega=\Omega_c$ with $\lambda_{12}=1$, for different mass splittings $m_2-m_1$. 
The light red region is ruled out due to the Higgs to invisible
constraint, and the region above the gold contour is ruled out in the SHP by the 
XENON1T data. The red solid lines denote the regions ruled out 
by the direct detection constraints on the assisted freeze out model.}
\label{fig:scan_1b}
\end{figure}

In Fig.~\ref{fig:scan_1b} we show some numerical values in the resonance region to exemplify 
the general behavior here. In this plot, we have considered contours with fixed mass
differences $m_2=m_1+1$ GeV (blue), $m_1-1$ GeV (green) and $m_1-5$ GeV (grey), which give
the observed DM relic abundance at $\lambda_{12} = 1$. 
As expected, in the normal mass hierarchy $S_1$ behaves very similar to $S$, however
in the inverted hierarchy, $\Omega_1$ drops enough such that direct detection constraint 
becomes effectively negligible, only ruling out parameter space outside the Higgs resonance region \footnote{As we pointed out in a footnote in page 7, near the resonance kinetic equilibrium can be lost, requiring a more precise treatment to obtain the correct relic abundance in this region. In our case, where only $S_1$ couples to the Higgs, this effect should be more prominent when $\Omega_1$ is dominant (normal hierarchy), with expected changes equivalent to those obtained in \cite{Binder:2017rgn} for the SHP. In the inverse hierarchy, where $\Omega_1$ is sub-dominant, we expect a small correction to the total relic abundance. We did not consider such a precise treatment in this work.}.
For comparison, we have also plotted contours where the SHP could account for the 100\%, 10\% and 1\% of the correct relic abundance with a black, blue and red dashed lines, respectively.
As expected, due to the additional degrees of freedom coming from $m_2$ and $\lambda_{12}$,
the two scalar model can explain the entire DM relic abundance in that region but also even at slightly smaller masses $m_1$. Thus, for masses near the Higgs resonance
the SHP parameter space opens up in the two scalar model and it remains to be seen,
whether WIMPs in the region can be fully ruled out as DM candidates.
\begin{figure}[h]
\centering\hspace*{-5mm}
\includegraphics[width=0.33\textwidth, trim=130 50 70 50, clip]{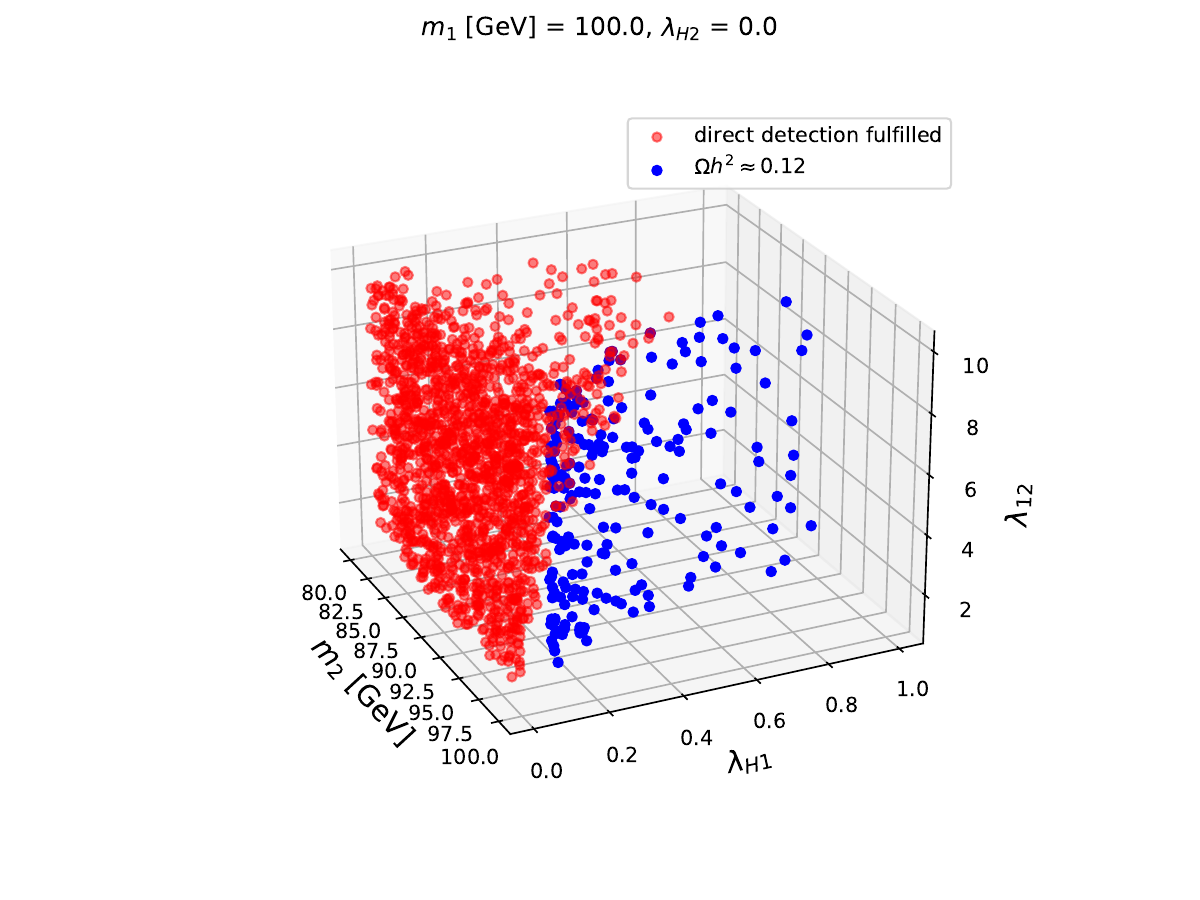}\hspace*{-2mm}
\includegraphics[width=0.33\textwidth, trim=10 5 40 50, clip]{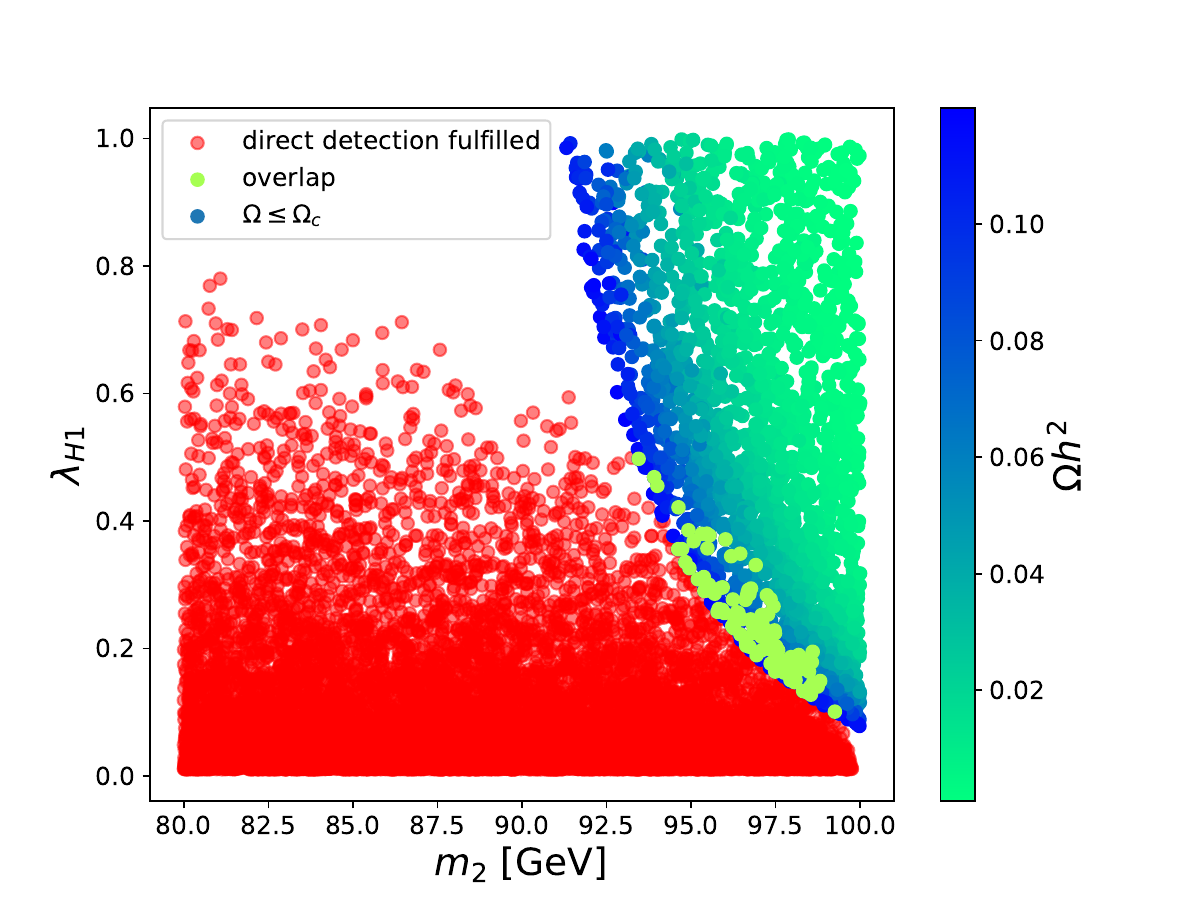}
\hspace*{-2mm}
\includegraphics[width=0.36\textwidth, trim=85 50 65 50, clip]{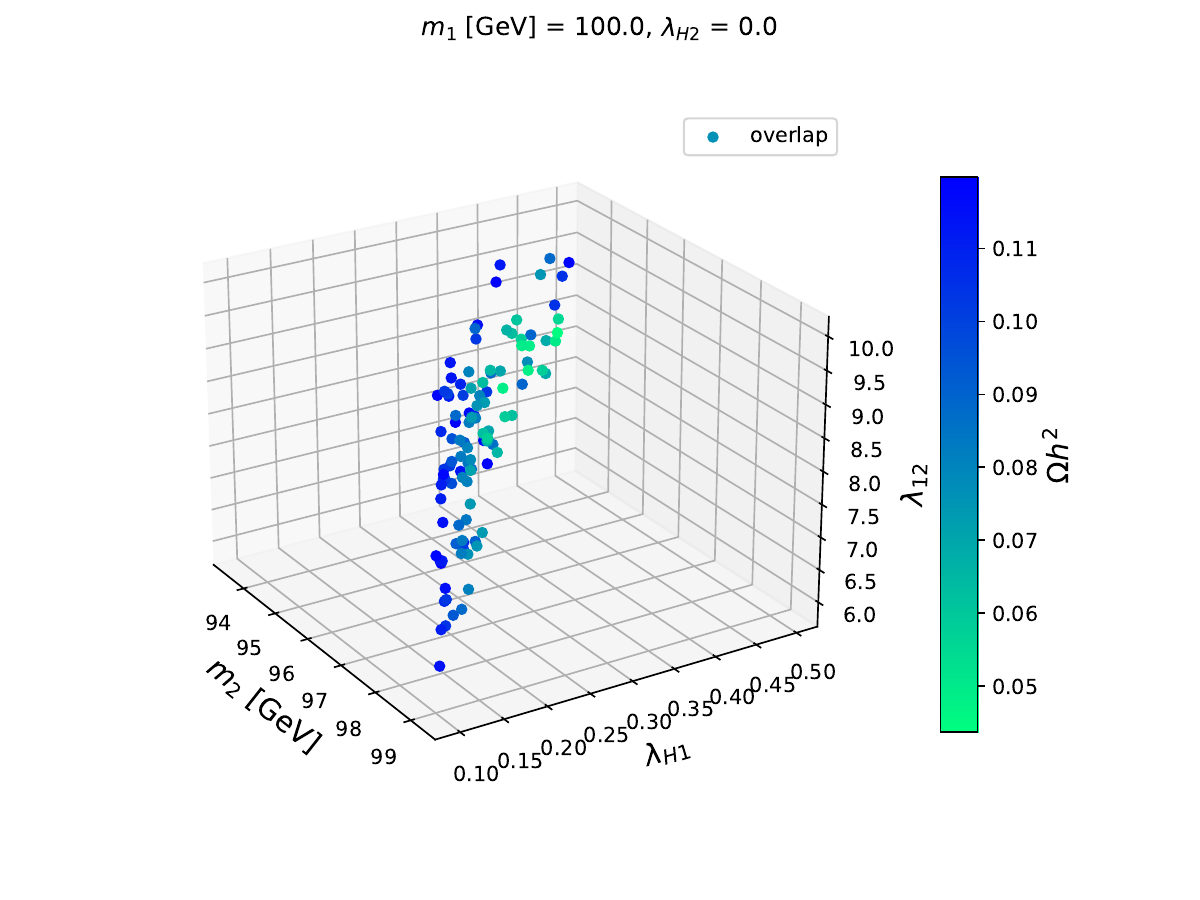}
\centering
\caption{Random scan on the inverted mass hierarchy ($m_2<m_1$) parameter space for fixed $m_1 = 100$ 
GeV and $\lambda_{H2}=0$.
The red points fulfill the direct detection constraints, while the blue-green points fulfill
$\Omega\leq\Omega_c$ (with $\Omega$ indicated by the color bar). The left plot shows points projected on the 3d parameter space $(m_1, \lambda_{H1}, \lambda_{12})$, the middle plot shows the projection on the plane $m_1-\lambda_{H1}$ plane, and the right plot shows just the overlap of points fulfilling the direct detection constraints and the relaxed abundance constraint $\Omega\leq\Omega_c$.}
\label{fig:scan_2a}
\end{figure}

Outside the Higgs resonance region, the most promising parameter space is in the inverted
mass hierarchy ($m_2 < m_1$), as by our analysis from section \ref{sec:assisted_freeze-out}, both the
abundance and direct detection constraints are most easily fulfilled there.
We thus ran larger scans of points $(m_2, \lambda_{H1}, \lambda_{12})$ for fixed $m_1=100$
GeV (Fig.~\ref{fig:scan_2a}) and $m_2<m_1$. Precisely, the parameter space is sampled using equal spacing for $\lambda_{12}\in(1, 10)$ and $m_2\in(80,100)$ and
 	logarithmic spacing for $\lambda_{H1}\in(0.01, 1)$ with a total number of points $n=18000$. The grid scan used equal spacing 
 	with 50 steps for $m_2\in(90,100)/(470, 500)$ and 20 steps for $\lambda_{H1}\in(0.05,0.6)/(0.3,2.5)$ and 
 	$\lambda_{12}\in(5,15)$.
In this region, points fulfilling the direct detection constraints (red) are either at very
small couplings $\lambda_{H1}$ and arbitrary $\lambda_{12}$, 
or, for larger couplings $\lambda_{H1}$ and $\lambda_{12}$ at masses $m_2$ mostly below that
fulfilling the abundance constraint for the given couplings.
Since in the inverted mass hierarchy a lower relic abundance at equal couplings can only
be achieved by a smaller mass difference, points fulfilling $\Omega<\Omega_c$ lie
opposite from the points fulfilling the direct detection constraints, with the points
fulfilling $\Omega=\Omega_c$ forming a boundary. Hence the overlap of these two regions
does not increase significantly when relaxing the abundance constraint. This fact is shown in Fig. \ref{fig:scan_2a} (middle) where the small overlap of points fulfilling the direct 
detection and relaxed abundance constraint is shown.
In particular from Fig. \ref{fig:scan_2a} (right) we see no overlap for 
$\lambda_{12}\lesssim 6$, since smaller $\lambda_{12}$ also requires small $\lambda_{H1}$ to pass the direct detection constraint which in turn produces a DM over-abundance.

\begin{figure}[t!]
\centering\hspace*{-5mm}
\includegraphics[width=0.55\textwidth]{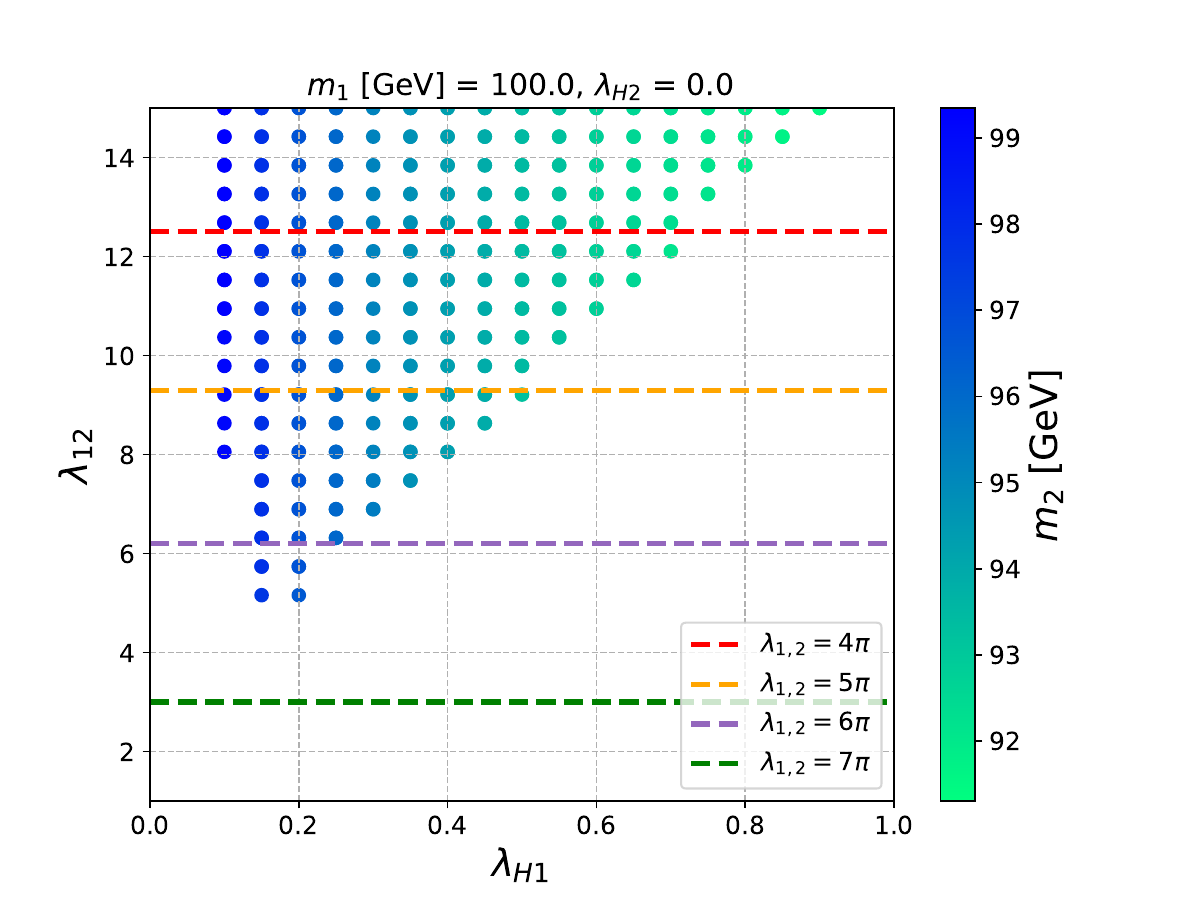}\hspace*{-5mm}
\includegraphics[width=0.55\textwidth]{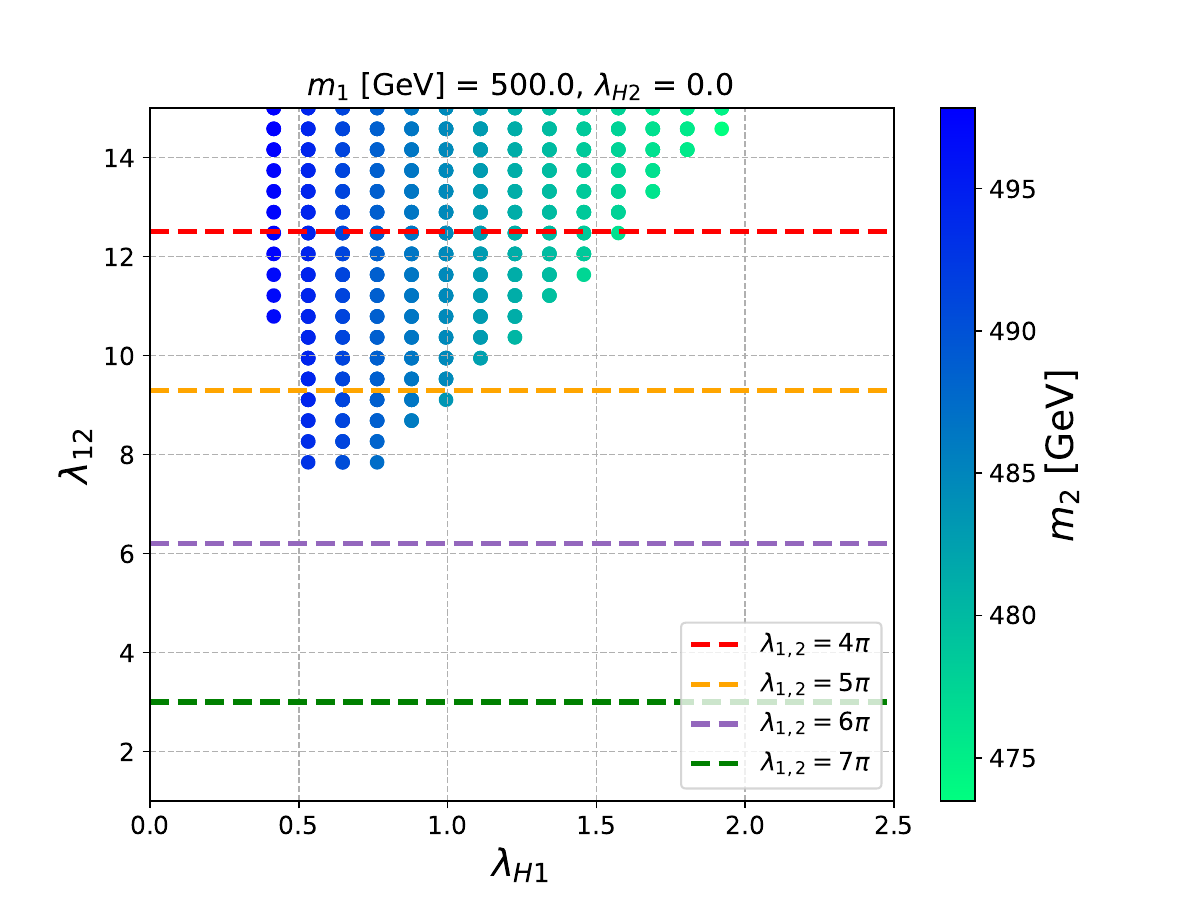}\par
\caption{Allowed points by experimental constraints considering $m_1 = 100$ GeV (left) and 500 GeV (right) projected in the plane $(\lambda_{H1},\lambda_{12})$. The color bar indicates the mass of $S_2$. The dashed lines correspond to the maximum value allowed for $\lambda_{12}$ considering that the theoretical constraints allowed quartic couplings up to the corresponding value indicated on the legend.}
\label{fig:scan_2b}
\end{figure}

To get a more precise picture of the interplay between the parameter space allowed by the experimental and the theoretical constraints, in Fig.~\ref{fig:scan_2b} we show a plot of the allowed points with $m_1 = 100$ and 500 GeV projected in the plane ($\lambda_{H1}, \lambda_{12}$), with a color bar for $m_2$. The left plot shows the same parameter space as Fig.~\ref{fig:scan_2a}, and confirms that sizable values for $\lambda_{12}$ are required in order to fulfill the experimental DM constraints. Further, as shown in the right plot, for a higher $m_1$, the smallest possible values for $\lambda_{12}$ increase as well. Interestingly, the smallest achievable $\lambda_{12}$ couplings in each case are around 6 and 8 (left/right plots), and they are achieved for very particular mass splittings, 1 GeV in the left case and around 5 GeV in the right plot. As suggested by Fig.~\ref{fig:scan_2b}, with growing $m_1$, the allowed parameter space shrinks. We have checked numerically that the minimum allowed $\lambda_{12}$ by direct detection and relic abundance constraints in the range for $m_1$ from $m_h/2$ up to 1 TeV does not change by more than a factor of two.

As was argued in Section \ref{modelandtheoconst}, it is necessary that the coupling values be in agreement with perturbativity and unitarity, which depend on the values of the quartic couplings. In the Fig.~\ref{fig:scan_2b} we show the contours of the unitarity constraints for different $\lambda_{1,2}$. The points below each dashed line are allowed by unitarity for a certain range of values of $\lambda_{1,2}$. For instance, all the points below the yellow dashed line are allowed for $\lambda_{1,2}$ up to $5\pi$. If the values of $\lambda_{1,2}$ are below $4\pi$, then the corresponding unitarity dashed line rises above the red dashed line, and the allowed parameter space will depend only on the perturbative limit assumed on $\lambda_{H1}, \lambda_{12}$. As a complementary analysis, in Appx.~\ref{oneloop} we discuss the effects of the radiative generation of $\lambda_{H2}$ on the results shown in Fig.~\ref{fig:scan_2b}.

\subsection{Landau Poles}
As a result of the previous section, high coupling values are necessary to obtain the correct relic abundance evading direct detection bounds, especially for $\lambda_{12}$. However, in a scalar theory like this one, such large couplings grow rapidly with the energy scale according to the one-loop renormalization group equations (RGE). Estimating to what extent this model will be valid in the high 
energy regime, i.e. before some of the couplings reach a \textit{Landau pole}, is therefore vital for its application. 
The criteria used here to find the Landau poles is simply one of the couplings exceeding $4\pi$ according to RGE evolution. In the following, we will study the energy validity of the model in a general way, then for simplicity assuming $\lambda_1$ and $\lambda_2$ negligible. 

Apart of the existing $\beta$ functions for the SM couplings, from the two DM scalar model we obtain \cite{Modak:2013jya}
\begin{eqnarray}
 16\pi^2\beta_{\lambda_{H}} &=&  \frac{3}{8}g_1^4 + \frac{9}{8}g_2^4 + \frac{3}{4}g_1^2 g_2^2 - 6y_t^4 + 24\lambda_H^2 + 12y_t^2 \lambda_H - 3g_1^2 \lambda_H - 9g_2^2 \lambda_H + \frac{1}{2}(\lambda_{H1}^2 + \lambda_{H2}^2) ,\nonumber\\
 16\pi^2\beta_{\lambda_{H1}}  &=&   \lambda_{H1}\left(12\lambda_H + \lambda_{12} + 4\lambda_{H1} + 6 y_t^2 -\frac{3}{2}g_1^2 - \frac{9}{2}g_2^2\right) + \lambda_{H2}\lambda_{12}, \nonumber\\
 16\pi^2\beta_{\lambda_{H2}}  &=&  \lambda_{H2}\left(12\lambda_H + \lambda_{12} + 4\lambda_{H2} + 6 y_t^2 -\frac{3}{2}g_1^2 - \frac{9}{2}g_2^2\right) + \lambda_{H1}\lambda_{12}, \nonumber\\
 16\pi^2\beta_{\lambda_{12}}  &=&   4\lambda_{12}^2 + 4\lambda_{H1}\lambda_{H2},
\end{eqnarray}\label{rge}
where $\beta_\lambda \equiv \mu \frac{d\lambda}{d\mu}$, with $\mu$ the renormalization scale, $g_1, g_2, g_3$ are the gauge couplings for $U(1), SU(2), SU(3)$, respectively, and $y_t$ is the top quark Yukawa coupling. We solve this system using \textit{RGE package}\footnote{RGErun 2: Solving Renormalization Group Equations in Effective Field Theories (K. Kannike).}, considering the following initial conditions evaluated at $\mu = m_t = 173$ GeV: $g_1(m_t) = \sqrt{5/3}\times 0.359, g_2(m_t) = 0.647, g_3(m_t) = 1.166, y_t(m_t) = 0.950, \lambda_H(m_t) = 0.128, \lambda_{H2}(m_t)=0$, whereas the values for $\lambda_{H1}(m_t)$ and $\lambda_{12}(m_t)$ are specified in the following discussion.
\begin{figure}[t!]
\centering\hspace*{-5mm}
\includegraphics[width=0.45\textwidth]{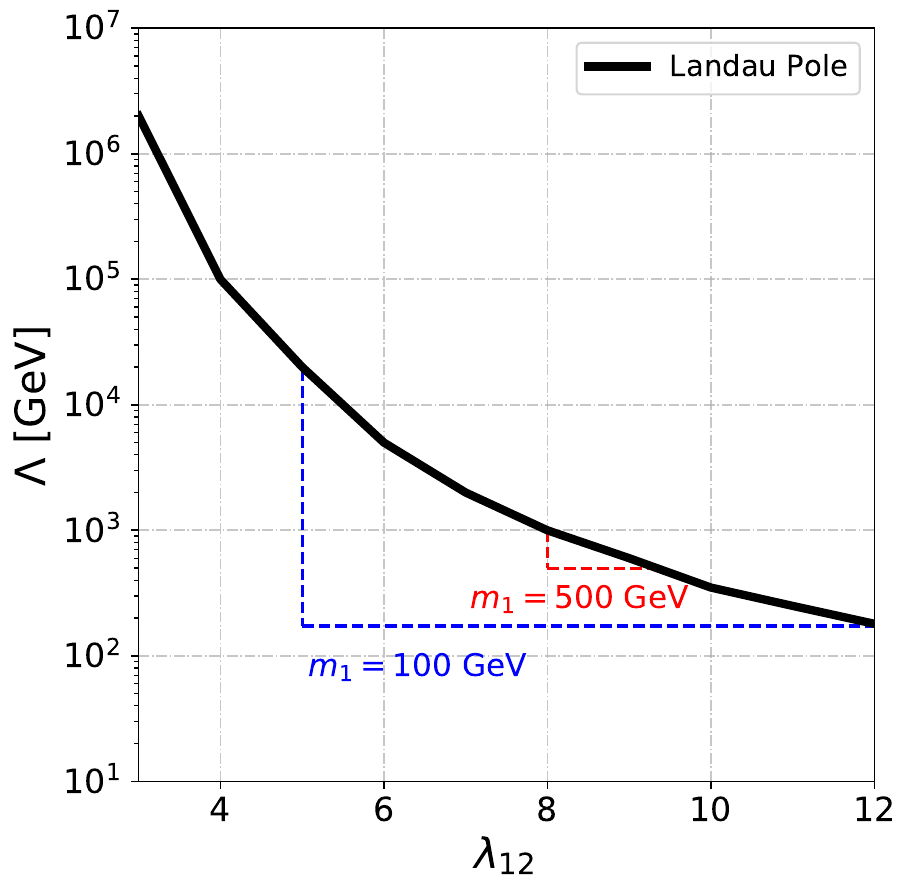}
\caption{Energy scale at which a Landau pole appears in the two-scalar DM assisted freeze-out framework as a function of $\lambda_{12}(m_t)$, considering $\lambda_{H1}(m_t) = 0.1, 0.5, 1.0$ and $1.5$. The regions enclosed by the dashed curves are allowed considering $m_1 = 100$ GeV (blue) and $m_2 = 500$ GeV (red). }
\label{fig:scan_3b}
\end{figure}
As the last line in eq.~\eqref{rge} shows $\beta_{\lambda_{12}}\propto \lambda_{12}^2$ and the values obtained in the previous section resulted to be $\lambda_{12}(m_t)\gtrsim 5$, a strongly growing behavior with energy is expected for $\lambda_{12}$, resulting in Landau poles at relatively low energy scales. In Fig.~\ref{fig:scan_3b} we show how the maximum energy $\Lambda$ at which the coupling $\lambda_{12}$ reaches its Landau pole (black solid line) decreases as $\lambda_{12}(m_t)$ becomes bigger. Variations of $\lambda_{H1}(m_t)$ in the range $0.1 - 1.5$ do not give rise to significant deviations respect to the black curve. Additionally, the parameter space found in the last section for $m_1 = 100$ and 500 GeV is reflected as the regions inside the dashed and continuous curves. That is, for $m_1 = 100$ GeV, the blue vertical dashed line is the minimum value found for $\lambda_{12}$, whereas the horizontal line is the initial renormalization scale $\mu = m_t$. As $\lambda_{12}$ at the EW scale gets larger, the validity of the model becomes greatly constrained at high energies, showing a Landau pole already at a few hundred GeV. For higher masses, the situation becomes even more drastic, as shown for $m_1 = 500$ GeV with the red dashed lines, where $\lambda_{12}\gtrsim 9$ reaches a Landau pole below the masses of the singlet scalars. 

Therefore, it is clear that high coupling values at the EW scale set limitations for the model at high energies, especially if we want to study the two-DM scalars in the assisted freeze-out in a cosmological context (e.g. \cite{Lerner:2009xg}). Possible extensions to this scenario can be thought of in order to evade such large couplings. For instance, adding to the present model a vector-like lepton which transform non-trivially under one of the discrete symmetries opens up a new annihilation channel for $S_1S_1$ annihilations, reducing significantly the high values for $\lambda_{12}$ required in the present work \cite{Saez:2021qta}. 

\section{Discussion and Summary}
In this paper we investigated to what extent the parameter space of the Singlet Higgs Portal can be reopened by
introducing a second DM candidate decoupled from the SM: \textit{Assisted freeze-out framework}. The outcome is positive. 



We conclude that, in order to evade direct detection and relic abundance constraints outside the Higgs resonance 
region it is required that one of the key new parameters $\lambda_{12}$ takes high values, which are constrained by theoretical considerations of the potential stability, perturbativity and unitarity. In this way small regions in the parameter space
reopen for DM masses around a few hundred GeV.
Additionally, our results are in accordance with what was found \cite{Maity:2019hre}, where the authors studied a model in an similar framework. In the present work, we explored in major detail the parameter space, concluding that outside the resonance region, DM can be explained in the two scalar model if $m_{decoupled}<m_{coupled}$ with a relative mass difference of up to
roughly 10\% in the most optimistic cases (i.e. if we allow a very large coupling $\lambda_{12}$).

Similarly to the SHP, the two real scalar model in the assisted freeze-out regime shows that parts of the Higgs resonance region remain not (yet) ruled out by experimental data. 
In fact, due to the additional degrees of freedom $m_2$ and $\lambda_{12}$, the two scalar model in the assisted freeze-out framework could completely explain DM for a broader range of values $m_1$ and $\lambda_{H1}$, however, the explanation would still require $m_1$ in the region of $m_h/2$ and simultaneously a very small mass splitting between $m_1$ and $m_2$.

Finally, we have shown that one-loop RGE imply that the high couplings at the EW scale found in this work, i.e. $\lambda_{12}\sim \mathcal{O}(5 - 10)$, imply a very low energy validity for the model with two stable scalars, requiring a UV completion at low energy scales for the model to be viable at higher energies. Alternatives have been proposed too, such as to include a vector-like fermion portal at the EW scale.

\section{Acknowledgment}
B.D.S. would like to thank Roger Galindo, Felipe Rojas, Sebastian Norero and Gorazd Cvetic for useful discussions, and to CONICYT Grant No. 74200120.

%% file: appendix.tex
\section{Average Annihilation Cross Sections}\label{App:AvgCs}
For a scattering process $ij\rightarrow ab$, the average cross section times velocity following a Maxwell-Boltzmann thermal distribution is given by
	\begin{align}
	\langle \sigma_{ijab} v \rangle = \dfrac{1}{n_{ie}n_{ie}}
	\int\dfrac{d^3p_i}{(2\pi)^3}\int\dfrac{d^3p_j}{(2\pi)^3}
	(\sigma_{ijab} v) \cdot e^{-(E_i+E_j)/T},
	\label{def_thermal_averaged_cross_section}	
	\end{align}
where $v$ denotes the M\o ller velocity, and $n_{ie}$ the equilibrium densities given in eq. \eqref{eqden}. The cross sections for the processes we are going to 
consider can be written as functions of $s:=(E_i+E_j)^2-(\vec{p_i}+\vec{p_j})^2$. 
The integrals in eq. \eqref{def_thermal_averaged_cross_section} then break down to just a 
single integral over s
	\begin{align}
	\langle \sigma v \rangle = \int_{(m_i+m_j)^2}^{\infty} ds
	\dfrac{(\sigma v)^{(cm)}\cdot s\cdot K_1(\tfrac{\sqrt{s}}{T})}{16Tm_i^2m_j^2
	K_2(\tfrac{m_i}{T})K_2(\tfrac{m_j}{T})}
	\sqrt{\dfrac{(m_i^2-m_j^2)^2}{s}+s-2(m_i^2+m_b^2)},
	\label{thermal_average_s_integral}	
	\end{align}
where $(\sigma v)^{(cm)}$ denotes the calculation in the center of mass frame and
$K_{\alpha}$ is the modified Bessel function of the second kind. 
Similarly, $\langle s\rangle$ (for $m_i=m_j$) can now be calculated in reverse. Since $s=4m_i^2+4p_{(cm)}^2$, then
	\begin{align}
	\langle s\rangle = \dfrac{1}{n_{ie}}\int\dfrac{d^3p}{(2\pi)^3}
	(4m_i^2+4p^2)e^{-E/T}=4m_i^2 + 12m_iTK_3(\tfrac{m_i}{T})/K_2(\tfrac{m_i}{T}).
	\end{align}
Away from resonances for large $x$, the averaged annihilation cross section is then approximately 
\begin{align}
	\langle\sigma v\rangle \approx \sigma v(\langle s\rangle)\approx\sigma v\left(4m^2_i(1+\tfrac{3}{x}\tfrac{\mu}{m_i})\right) \approx 
	\sigma v\left(4m^2_i\right)
\end{align}

\begin{figure}[H]
	\centering 
	\includegraphics[width=0.49\textwidth, trim=5 0 20 0, clip]{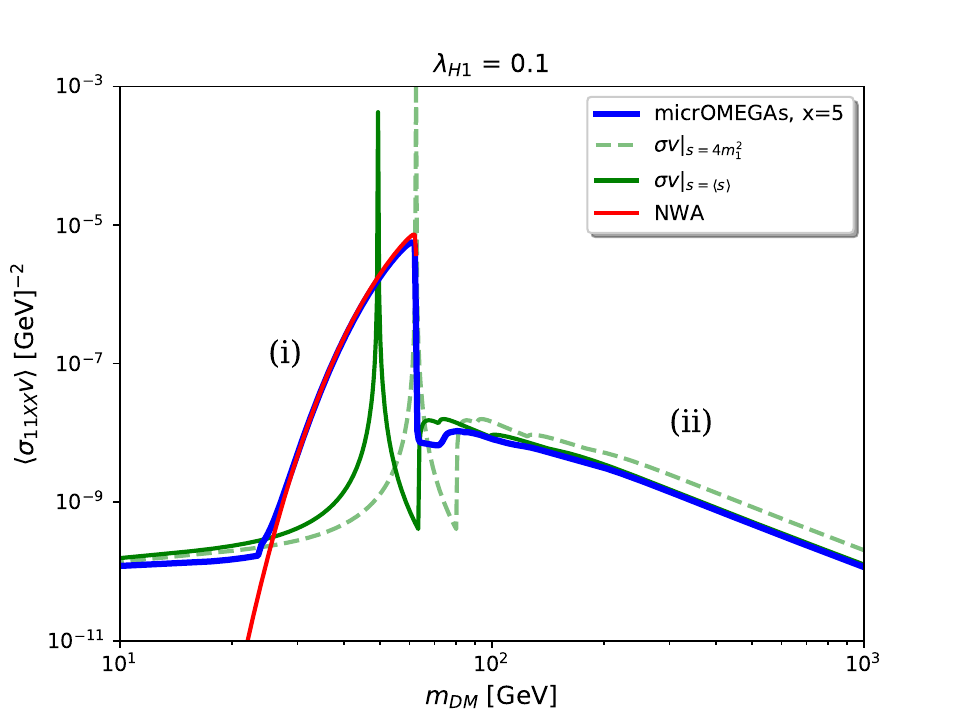}\hspace*{-2mm}
	\includegraphics[width=0.49\textwidth, trim=5 0 20 0, clip]{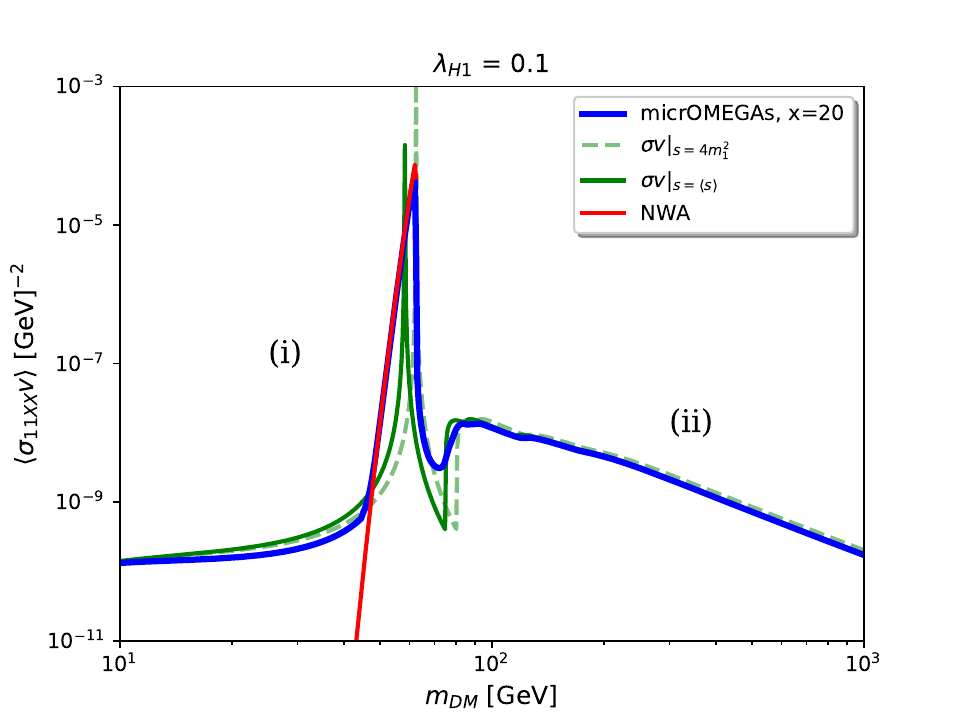}
	\caption[Caption for LOF]{Comparison of $\langle\sigma_{11XX} v\rangle$ as a function of $m_1$ calculated using the NWA (red) and
		the constant approximation with (green) and without the $\frac{3}{x}$ correction (green dashed). We set $\lambda_{H1}=0.1$
		and $x=5$ (left), respectively $x=20$ (right). For comparison we also plotted the result obtained with $\micrOMEGAs$ (blue solid line).}
	\label{fig:avrg_apprx_comparison}
\end{figure}

Near the Higgs resonance, i.e. $m_1 \lesssim m_h/2$, we used the well known approximation for $\sigma v$ in eq.~(\ref{thermal_average_s_integral}), which replaced the propagator by a Dirac delta function 
		(Narrow Width Approximation):
		\begin{align}\label{nwa}
		\dfrac{1}{(s-m_h^2)^2+m_h^2\Gamma^2} \approx \dfrac{\pi}{m_h\Gamma}
		\delta(s-m_h^2),
		\end{align}		
		where $\Gamma$ is the total Higgs decay width, including the invisible decays into
		DM particles. Using this approximation, the integral in eq.~(\ref{thermal_average_s_integral}) becomes an algebraic expression.
In Fig.~\ref{fig:avrg_apprx_comparison}  we compare the approximations to the result from $\micrOMEGAs$ in the region (i)
below and (ii) above the Higgs resonance. Considering two temperatures, $x = 5$ and $20$, we show that our approximation (red solid curve) matches very well the $\micrOMEGAs$ result in the resonance region (i).

%% file: appendixb.tex
\section{$S_2S_2h$ vertex at one-loop}\label{oneloop}
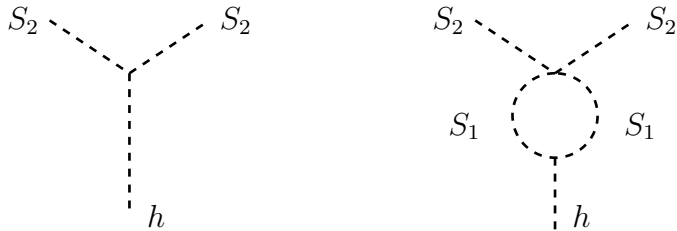
\begin{figure}[H]
\centering
\begin{tikzpicture}[line width=1.0 pt, scale=0.7]
\begin{scope}[shift={(0,0)}]
	\draw[scalarnoarrow](-2.5,1) -- (-1,0);
	\draw[scalarnoarrow](-1,0) -- (0.5,1);
	\draw[scalarnoarrow](-1,0) -- (-1,-2.7);
	\node at (-3,1.0) {$S_2$};
	\node at (1,1.0) {$S_2$};
    \node at (-0.5,-2.7) {$h$};
\end{scope}

\begin{scope}[shift={(8,0)}]
	\draw[scalarnoarrow](-2.5,1) -- (-1,0);
	\draw[scalarnoarrow](-1,0) -- (0.5,1);
	\draw[scalarnoarrow](-1,0) arc (-270:90:0.8);
	\draw[scalarnoarrow](-1,-1.6) -- (-1,-3);
	\node at (-3,1.0) {$S_2$};
	\node at (1,1.0) {$S_2$};
	\node at (-2.7,-1) {$S_1$};
    \node at (0.6,-1) {$S_1$};
    \node at (-0.5,-2.7) {$h$};
\end{scope}
\end{tikzpicture}
\caption{\footnotesize Tree-level $S_2S_2h$ vertex and the main one-loop corrections in the two real scalar model with $Z_2\times Z_2'$ symmetry in the limit $\lambda_{H2}\ll 1$.} \label{diagrams}
\end{figure}
In this appendix we show the calculation of the one-loop $\lambda_{H_2}$ coupling and its effects on the relic abundances of $S_1$ and $S_2$. As we are assuming that at tree level $\lambda_{H2}\ll (\lambda_{H1}, \lambda_{12})$, the only diagram which contributes to the vertex $S_2S_2h$ at one-loop is given in the right side of Fig.~\ref{diagrams}, whose amplitude is given by
\begin{eqnarray}
 \Gamma^\text{1-loop}(p^2) = (-i\lambda_{12})(-i\lambda_{H1}v_H)\frac{1}{2}\int \frac{d^4p_1}{(2\pi)^4}\frac{i}{(p_1 - p)^2 - m_1^2}\frac{i}{p_1^2 - m_1^2}, 
\end{eqnarray}
where $p_1 + p_2 = p$ and the $1/2$ is a symmetry factor due to the scalar loop. After some manipulation via Feynman parameters, and using dimensional regularization, we have 
\begin{eqnarray}\label{dimreg}
 \Gamma^\text{1-loop}(p^2)  = \frac{1}{16\pi^2}\lambda_{12}\lambda_{H1}B_0(p^2,m_1^2,m_1^2), 
\end{eqnarray}
where $B_0(p^2,m_1^2,m_1^2) = 2/\epsilon + \mathcal{B}(p^2,m_1^2,m_1^2)$, with
 \begin{eqnarray}
 \mathcal{B}(p^2,m_1^2,m_1^2) = -\log\frac{m_1^2}{\tilde{\mu}^2} - \int_0^1 dx \log\left[1 - x(1-x)\frac{p^2}{m_1^2}\right]
\end{eqnarray}
where $\tilde{\mu}^2 =4\pi e^{-\gamma_E}\mu^2$ is a mass dimension parameter introduced to keep coupling constants dimensionless in $d$-dimensions. 

The renormalized (finite) amplitude is given by
\begin{eqnarray}\label{masterGamma}
 \Gamma(p^2) &=& \Gamma^\text{tree} + \Gamma^\text{1-loop} + \Gamma^\text{counter.} \nonumber \\
             &=& \lambda_{H2}v_H + \frac{1}{16\pi^2}\lambda_{12}\lambda_{H1}v_HB_0(p^2,m_1^2,m_1^2) + \delta_2 \lambda_{H2}v_H
\end{eqnarray}
where $\delta_{2}$ corresponds to the counter-term which is going to cancel the divergence from the second term in the right side of eq. \eqref{masterGamma}. After imposing the renormalization condition $\Gamma (p^2_0) = \lambda_{H2}v_H$, with $p_0^2$ some low energy scale, the amplitude becomes: 
\begin{eqnarray}\label{integral}
 \Gamma(p^2) = \lambda_{H2}v_H + \frac{1}{16\pi^2}\lambda_{12}\lambda_{H1}v_H\int_0^1 dx \log\left[\frac{1 - x(1-x)\frac{p^2_0}{m_1^2}}{1 - x(1-x)\frac{p^2}{m_1^2}}\right]
\end{eqnarray}
Taking that the renormalization scale very low (i.e. $p^2_0 \ll (m_1^2, m_2^2)$), and considering that at freeze-out we have that $p^2 \approx 4m_2^2$, then eq. \eqref{integral} becomes
\begin{eqnarray}\label{lh2rad}
  \Gamma_{S_2S_2h}  \approx \left(\lambda_{H2} + \frac{\lambda_{12}\lambda_{H1}}{16\pi^2}\right)v_H.
\end{eqnarray}
for $m_2 \lesssim m_1$, in agreement with \cite{Casas:2017jjg}. Additionally, the loop contribution has a completely negligible effect for direct detection such as XENON1T, due to the low transfer momentum $|p| \sim \mathcal{O}(\text{keV})$.

\begin{figure}[h!]
\centering
\includegraphics[width=.45\textwidth]{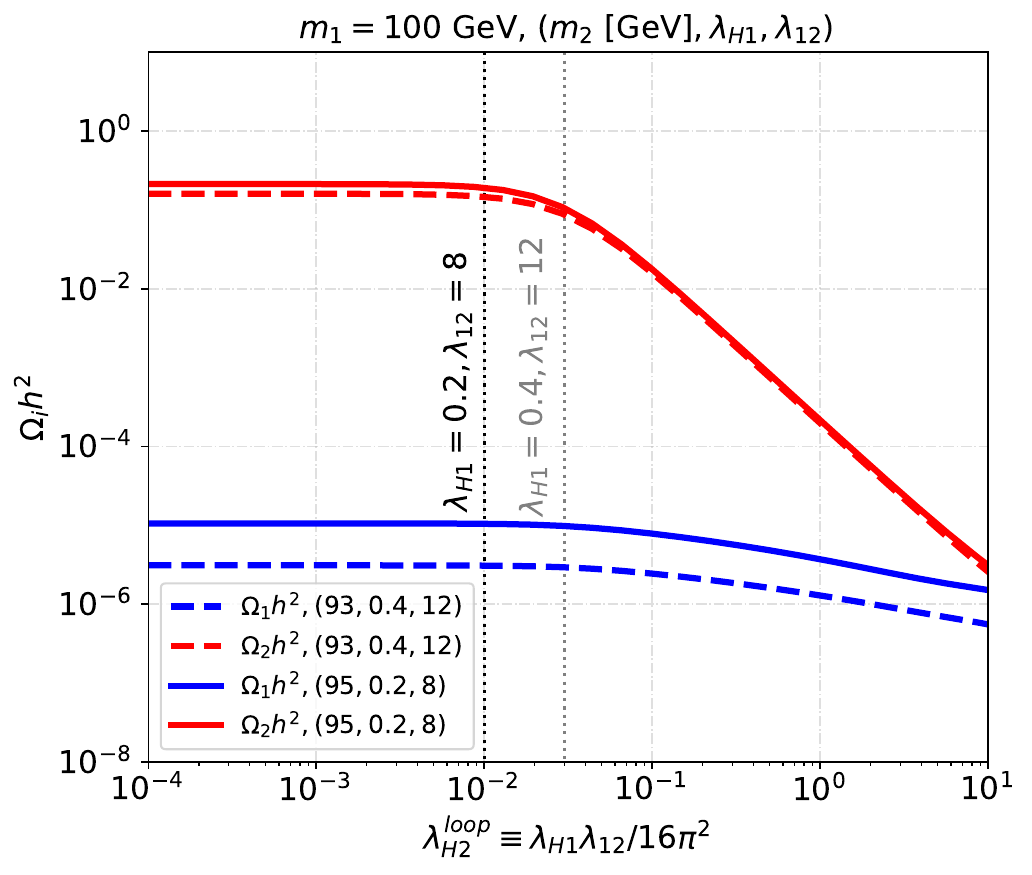}\quad\quad
\includegraphics[width=.45\textwidth]{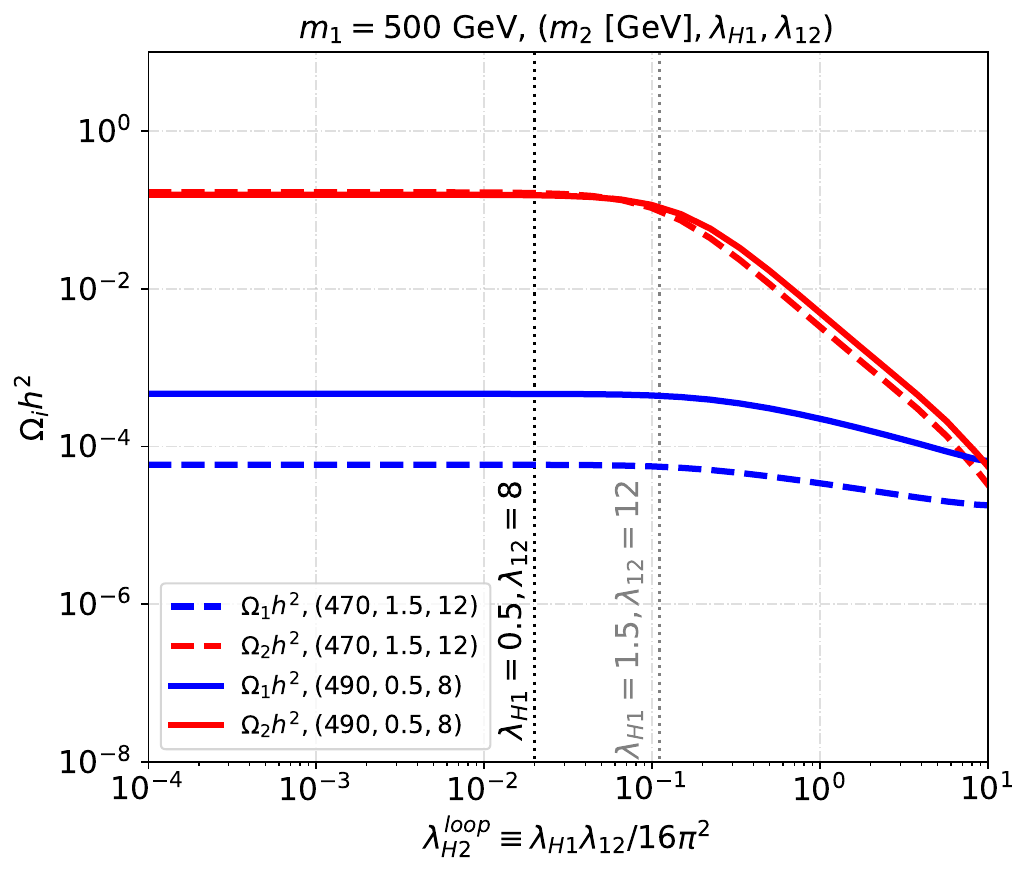}
\caption{Relic abundances as a function of $\lambda_{H2}\equiv \lambda_{H1}\lambda_{12}/16\pi^2$, for $m_1 = 100$ GeV (left) and $m_1 = 500$ GeV (right), with the rest of the parameters specified in the plots.}
\label{fig:radiative_ome}
\end{figure}

In Fig.~\ref{fig:radiative_ome} we show the changes in the relic abundances as a function of $\lambda_{H2}$, considering some points of Fig.~\ref{fig:scan_2b} with the correct relic abundance and evading direct detection. The vertical lines represent the one-loop value of $\lambda_{H2}$ ($=\lambda_{H1}\lambda_{12}/16\pi^2$) for the chosen points indicated in the plot. In both plots, the highest combination of values of $\lambda_{H1}\lambda_{12}$ (grey dashed lines) deviate the relic abundances predictions in a factor of two with respect to the simplest case in which $\lambda_{H2}= 0$. Note that these high coupling values must to be in agreement with the chosen perturbative/unitarity criteria. As it is shown in Fig.~\ref{fig:scan_2b}, the values for the pair ($\lambda_{H1},\lambda_{12}$) given by the dashed vertical lines in both plots of Fig.~\ref{fig:radiative_ome} are valid provided that $(\lambda_1,\lambda_2) \leq 4\pi$. Oppositely, higher values for $(\lambda_1,\lambda_2)$ implies lower values for the pair ($\lambda_{H1},\lambda_{12}$), then the effect of the radiative coupling $\lambda_{H2}$ becomes weaker. Finally, we have checked that these typical radiative values for $\lambda_{H2}$ shown in Fig.~\ref{fig:radiative_ome} do not change our conclusions in Sec.~\ref{modelandtheoconst}.

